\documentclass[aps,prc,twocolumn,superscriptaddress,nofootinbib,floatfix,preprintnumbers,10pt]{revtex4-2}

\usepackage{newtxtext,newtxmath}  

\usepackage{amsmath,amssymb,amsfonts,bm,mathtools}

\usepackage{euscript}

\usepackage{graphicx}
\usepackage{color}
\usepackage{booktabs,tabulary}
\usepackage{pstricks,pst-node,pst-text,pst-3d}  
\usepackage{comment}

\usepackage{relsize}
\usepackage{dsfont}
\usepackage[version=4]{mhchem}
\usepackage[makeroom]{cancel}
\usepackage{physics}
\usepackage{siunitx}
\AtBeginDocument{\RenewCommandCopy\qty\SI}
\usepackage{nicefrac,xfrac}

\usepackage[colorlinks=true, pdfstartview=FitV, linkcolor=blue, citecolor=blue, urlcolor=blue]{hyperref}

\newcommand*\diff{\mathop{}\!\mathrm{d}}

\newcommand{\be}{\begin{equation}}
	\newcommand{\ee}{\end{equation}}

\begin{document}
	
	\newcommand{\IdentityMat}{\mathbb{1}}
	\newcommand{\varQ}{\mathbf{q}}
	\newcommand{\varK}{\mathbf{k}}
	\newcommand{\varX}{\mathbf{x}}
	\newcommand{\varR}{\bm r}
	\newcommand{\Sch}{ Schr\"{o}dinger }
	\newcommand{\coeffSch}{-\frac{\hbar^2}{2m}}
	\newcommand{\etal}{\textit{et al.} }
	\newcommand{\secondfdv}[3]{ \frac{\delta^2 {#1}}{\delta #2 \, \delta #3} }
	
	\newcommand{\FM}[1]{{\color{magenta} #1}}
	\newcommand{\slava}[1]{{\color{blue} #1}}

	\title{Bridging reaction theory and nuclear structure in $\pi^\pm$-${}^{48}$Ca scattering}
	
	\author{Viacheslav Tsaran}
	\email{vitsaran@uni-mainz.de}
	\affiliation{Institut f\"{u}r Kernphysik and PRISMA+ Cluster of Excellence, Johannes Gutenberg-Universit\"{a}t Mainz, 55128 Mainz, Germany}

	\author{Francesco Marino}
	\email{frmarino@uni-mainz.de}
	\affiliation{Institut f\"{u}r Kernphysik and PRISMA+ Cluster of Excellence, Johannes Gutenberg-Universit\"{a}t Mainz, 55128 Mainz, Germany}
	
	\author{Sonia Bacca}
	\email{s.bacca@uni-mainz.de}
	\affiliation{Institut f\"{u}r Kernphysik and PRISMA+ Cluster of Excellence, Johannes Gutenberg-Universit\"{a}t Mainz, 55128 Mainz, Germany}
	\affiliation{Helmholtz-Institut Mainz, Johannes Gutenberg-Universität Mainz, D-55099 Mainz, Germany}
	
	\author{Francesca Bonaiti}
	\email{bonaiti@frib.msu.edu}
	\affiliation{ Facility for Rare Isotope Beams, Michigan State University, East Lansing, MI 48824, USA}
	\affiliation{Physics Division, Oak Ridge National Laboratory, Oak Ridge, TN 37831, USA}
	
	\author{Marc Vanderhaeghen}
	\email{vandma00@uni-mainz.de}
	\affiliation{Institut f\"{u}r Kernphysik and PRISMA+ Cluster of Excellence, Johannes Gutenberg-Universit\"{a}t Mainz, 55128 Mainz, Germany}
	\affiliation{Helmholtz-Institut Mainz, Johannes Gutenberg-Universität Mainz, D-55099 Mainz, Germany}
	
	\date{\today}
	
	\begin{abstract}
		We extend the pion-nucleus multiple-scattering framework to include detailed second-order rescattering dynamics for nuclei with non-zero isospin. 
		To account for intermediate charge-exchange and nucleon spin-flip effects, we develop a scattering potential that depends on the one- and two-body densities of the target nucleus. 
		We compute one-body densities from coupled-cluster theory and two-body densities within the Hartree-Fock approximation.
		To estimate theoretical uncertainties, we employ modern nuclear Hamiltonians derived from chiral effective field theory. 
		While the sensitivity to nuclear structure details is mild, second-order corrections are found to be sizeable and essential for accurately reproducing differential cross sections measured in $\pi^\pm$-${}^{48}$Ca elastic scattering within the $\Delta(1232)$-resonance region.
	\end{abstract}
	
	\maketitle
	
	\section{Introduction}
	Despite its long history~\cite{Ericson:1988gk, Johnson:1992zw}, the study of pion-nucleus interactions remains highly relevant in modern nuclear physics~\cite{PinzonGuerra:2018rju, LArIAT:2021yix}. 
	Pion-nucleus interactions play an important role in modern long-baseline neutrino oscillation experiments~\cite{DUNE:2016hlj, MicroBooNE:2016pwy, Hyper-KamiokandeProto-:2015xww, NOvA:2007rmc}. At neutrino energies below approximately \SI{1.5}{GeV}, pion production in neutrino-nucleus collisions is the dominant resonant process, significantly contributing to inclusive and semi-inclusive neutrino-nucleus scattering cross sections~\cite{Formaggio:2012cpf}. 
	Achieving the precision goal of a few percent level in these measurements requires accurate modeling of neutrino-nucleus interactions, including a reliable description of pion-nucleus final-state interactions such as rescattering, charge exchange, and absorption within the detector material~\cite{PinzonGuerra:2018rju,LArIAT:2021yix}. Relevant targets in this respect are $^{12}$C, $^{16}$O, and  $^{40}$Ar, the first two nuclei having the same number of protons and neutrons, as opposed to the latter.
	Consequently, studying pion scattering off nuclei with non-zero isospin is valuable for improving pion-nucleus interaction models employed by DUNE~\cite{DUNE:2016hlj} and MicroBooNE~\cite{MicroBooNE:2016pwy} experiments, which use ${}^{40}$Ar-based time projection chambers.

	Pion-nucleus interactions may also be used to extract the neutron skin thickness of finite nuclei with the goal of improving our understanding of the nuclear equation of state~\cite{Johnson:1978hz, Tarbert:2013jze, Thiel:2019tkm}.
	Coherent $\pi^0$ photoproduction on ${}^{48}$Ca is presently being pursued by the A2@MAMI Collaboration~\cite{Pedroni:2016, Watts:EPIC,WattsPC}. In this context, a reliable theoretical description of pion scattering off ${}^{48}$Ca, a non-zero isospin nucleus, is essential for modeling nuclear coherent pion photoproduction.
	Furthermore, recent calculations of $\pi^0$ coherent photoproduction below the $\Delta(1232)$ resonance energy within a distorted wave impulse approximation model, considering only a one-body photoproduction mechanism in the nucleus, have shown a low sensitivity to variations of the neutron skin thickness~\cite{Colomer:2022nqi}. To better quantify the model dependence of this reaction and its sensitivity to the neutron skin, it is of utmost importance to improve the formulation of the underlying scattering theory.
	
	In this paper, we study $\pi^\pm$-$^{48}$Ca scattering to set the stage towards a complete modeling of coherent pion photoproduction off $^{48}$Ca. We build on previous work on pion scattering off spin-isospin-zero nuclei, as developed in Ref.~\cite{Tsaran:2023qjx}.
	Within that framework, a key component is the second-order scattering potential, which incorporates a detailed treatment of pion rescattering on a nucleon pair with an intermediate nuclear excitation.
	To account for in-medium modifications of the elementary pion-nucleon scattering amplitudes, three parameters were introduced and fitted to $\pi^\pm$-${}^{12}$C scattering data in the pion laboratory kinetic energy range of 80--\SI{180}{MeV}, where the cross sections are particularly sensitive to the $\Delta$-resonance properties.
	The resulting potential, with no additional tuning, successfully reproduced $\pi^\pm$ scattering data for ${}^{16}$O, ${}^{28}$Si, and ${}^{40}$Ca~\cite{Tsaran:2023qjx}. 
	Furthermore, the same approach and effective $\Delta$ self-energy were applied in~\cite{Tsaran:2024sue} to successfully describe coherent $\pi^0$ photoproduction on ${}^{12}$C and ${}^{40}$Ca, reinforcing the consistency and universality of the approach for nuclei with zero spin and isospin.
	
	Here, we extend this reaction theory to nuclear targets with non-zero isospin. For this purpose, we generalize the derivation of the second-order pion-nucleus potential within the framework of multiple-scattering theory~\cite{Tsaran:2023qjx}.
	The resulting potential depends on several nuclear structure inputs, such as the proton and neutron one- and two-body densities. We compute one-body densities
	employing coupled-cluster theory~\cite{Hagen2014Review, Hagen_2016_rev, ShavittBartlett} and  two-body densities in the Hartree-Fock approximation. 
	Starting from various nuclear Hamiltonians derived from chiral effective field theory~\cite{Epelbaum2009, machleidt2011, hammer2020}, we investigate the sensitivity of scattering observables to the underlying nuclear structure. This work contributes to an ongoing effort to bridge nuclear structure and reaction theories, see for example, Refs.~\cite{Vorabbi2016, Rotureau2018, Idini:2019hkq, Hebborn:2022vzm, Vorabbi:2023mml}, where first-order scattering theory combined with one-body densities has been successfully applied to nucleon-nucleus cross sections.
	Here, we tackle for the first time the second-order scattering theory, albeit in the context of pion scattering. 
	
	The paper is organized as follows. 
	In Section~\ref{sec:multiple-scattering}, we review the key elements of the second-order pion-nucleus scattering potential formalism and extend it to nuclei with non-zero isospin. 
	Section~\ref{sec: nuclear structure} details the nuclear structure quantities, namely one-body densities and two-nucleon correlation functions, that enter the developed potential. 
	In Section~\ref{sec:fits-results}, we present a comparison between our model predictions and the experimental data for $\pi^\pm$-${}^{48}$Ca elastic scattering. 
	Finally, we give our conclusions and outlook in Section~\ref{sec:conclusion}.
	
	\section{Multiple-scattering formalism}
	\label{sec:multiple-scattering}
	
	The hadronic component of the amplitude describing pion-nucleus elastic scattering is given by the scattering amplitude $F$, which encapsulates the interactions between the incoming pion and the target nucleus. 
	These interactions include multiple scattering processes, resonance formations, and nuclear-medium effects such as pion absorption. 
	In our approach, we build upon the Kerman-McManus-Thaler formulation of multiple-scattering theory~\cite{Kerman:1959fr}, where  the elastic scattering amplitude $F$ is found as  solution of the Lippmann-Schwinger-type integral equation
	\begin{multline}
		F(\bm k^\prime, \bm k) = V(\bm k^\prime, \bm k) \\
		- \frac{A - 1}A  \int \frac{\diff \bm k^{\prime\prime}}{2\pi^2} \frac{V(\bm k', \bm k'') F(\bm k'', \bm k)}{k_0^2 - {k''}^2 + i \, \varepsilon},
		\label{LSh-pseudo-classical}
	\end{multline}
	where $\bm k$  and $\bm k'$ are the initial and final pion momenta in pion-nucleus center-of-mass (c.m.) frame, respectively, $k_0$ is the on-shell pion momentum, $A$ is the number of nucleons in the target nucleus, and we use the convention $\hbar = c = 1$.
	The solution of Eq.~(\ref{LSh-pseudo-classical}) accounts for pion rescattering on the nucleus in its ground state (g.s.) to all orders in perturbation theory, while the dynamics of nuclear excitation is encapsulated within the scattering potential $V$.
	In this way, the pion-nucleus scattering potential incorporates contributions from all possible intermediate excited states of the target nucleus.
	
	In our approach, we approximate $V$ using the first two terms of the corresponding scattering series, leading to~\cite{Tsaran:2023qjx}
	\begin{subequations}
		\begin{align}
			&V(\bm k^\prime, \bm k) \approx V^{[1]}(\bm k^\prime, \bm k) + V^{[2]}(\bm k^\prime, \bm k)
			\label{V-2term-approx}
			\\
			&V^{[1]}(\bm k', \bm k) = - A \frac{\sqrt{\mathscr{M}(k') \mathscr{M}(k)}}{2\pi} \langle \pi(\bm k'), \Psi_0 | \hat {\scriptstyle \mathcal{T}} | \pi(\bm k), \Psi_0 \rangle,
			\label{V_1st-def}
			\\
			&V^{[2]}(\bm k', \bm k) = - A(A-1) \frac{\sqrt{\mathscr{M}(k') \mathscr{M}(k)}}{2\pi} \notag
			\\
			& \qquad\qquad\qquad\ \ \ \times
			\langle \pi(\bm k'), \Psi_0 | \hat {\scriptstyle \mathcal{T}}_2  \hat G \hat P_\emptyset \hat {\scriptstyle \mathcal{T}}_1 | \pi(\bm k), \Psi_0 \rangle,
			\label{V_2nd-def}
		\end{align}%
		\label{V-all-def}%
	\end{subequations}
	where $|\Psi_0 \rangle$ is the g.s. wave function of the target nucleus and $\hat {\scriptstyle \mathcal{T}}_i$ represents the transition amplitude operator for pion scattering on the $i$th nucleon within the nucleus. 
	For simplicity, the subscript will be omitted when it is not necessary.
	The projection operator $\hat P_\emptyset =  \hat{ \mathds{1} } - |\Psi_0 \rangle \langle \Psi_0|$ ensures that intermediate states exclude the nuclear g.s., and $\hat G$ is the pion-nucleus Green's function.
	Here, $\mathscr{M}$ denotes the off-shell analog of the relativistic reduced mass of the pion-nucleus system in the pion-nucleus c.m. frame, which is given by
	\be
	\mathscr{M}(k) = \frac{[E(k_0) + E(k)][\omega(k_0)E_A(k_0) + \omega(k)E_A(k)]}{2 \left(E^2(k_0) + E^2(k)\right)},
	\ee
	where $\omega(k)$  and $E_A(k)$ are the relativistic energy of the pion and the nucleus, respectively, as functions of the c.m. momentum $k = \abs{\bm k}$, and $E(k) = \omega(k) + E_A(k)$.
	
	\begin{figure}[!tb]
		\center{\includegraphics[width=0.8\linewidth]{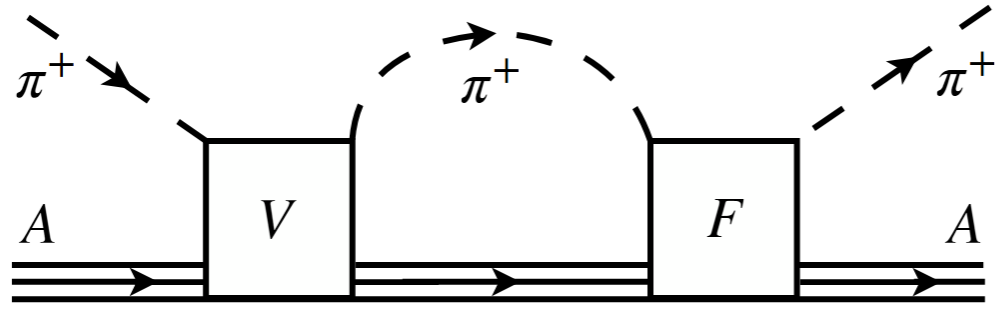}}
		\center{\includegraphics[width=0.8\linewidth]{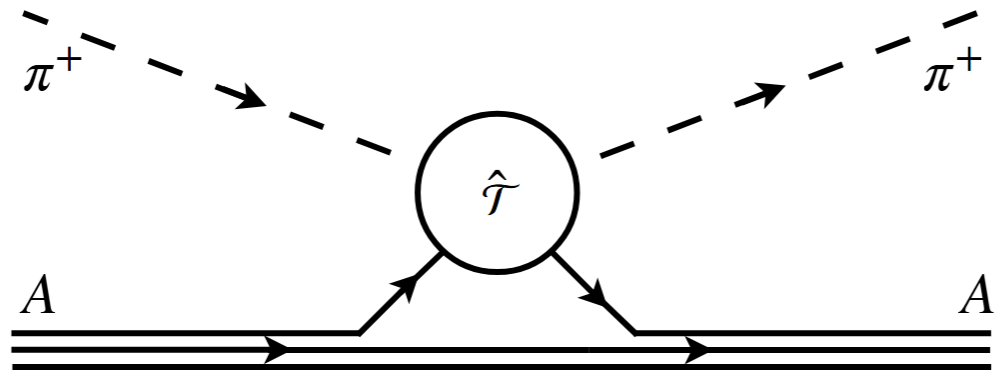}}
		\center{\includegraphics[width=0.8\linewidth]{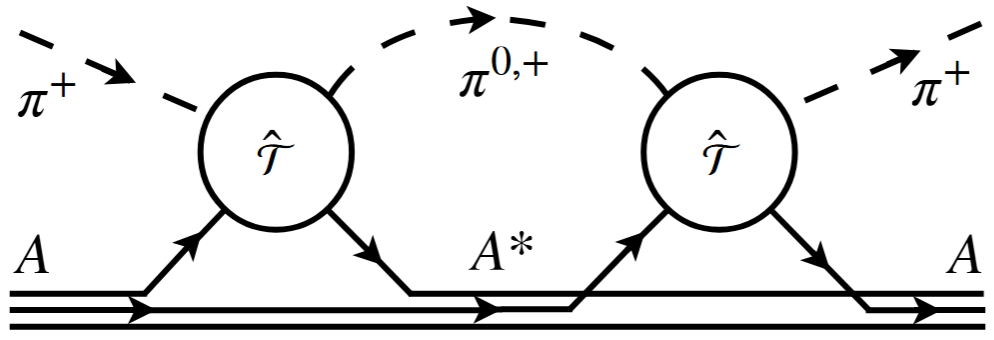}}
		\caption{
			Diagrammatic representation of the multiple-scattering formalism for $\pi^+$-nucleus scattering.
			The upper panel represents the second term on the right-hand side of Eq.~(\ref{LSh-pseudo-classical}).
			The middle and lower panels depict the first and second-order parts of the scattering potential $V$, respectively.
			$A$ and $A^*$ denote the nucleus in its ground and excited states, respectively.
		}
		\label{fig::T-diagrams}
	\end{figure}

	Figure~\ref{fig::T-diagrams} illustrates pion-nucleus scattering within the multiple-scattering formalism. 
	The upper panel depicts the all-order rescattering of the outgoing pion off the nucleus in its ground state, corresponding to the second term on the right-hand side of Eq.~(\ref{LSh-pseudo-classical}).
	The middle and lower panels depict the contributions of first and second order to the scattering potential $V$, corresponding to Eqs.~(\ref{V_1st-def}) and~(\ref{V_2nd-def}), respectively. 
	The lower panel specifically illustrates the second-order contribution, where the pion first scatters off one nucleon, exciting the nucleus, then propagates and undergoes a second scattering event that de-excites the nucleus.
	This two-nucleon mechanism facilitates charge exchange and/or spin flip between nucleons via the intermediate pion, ultimately bringing the nucleus back to its ground state.

	To evaluate the matrix elements in Eqs.~(\ref{V-all-def}), we employ the \textit{impulse approximation}~\cite{Ernst:1974hi}, 
	assuming that $\hat {\scriptstyle \mathcal{T}}$ is related to the pion-nucleon scattering amplitude in the pion-nucleon c.m. frame as
	\begin{multline}
		\langle \pi(\bm k^\prime), N(\bm p^\prime) | \hat {\scriptstyle \mathcal{T}} | \pi(\bm k), N(\bm p) \rangle  \approx { - (2\pi)^4}
		\delta(\bm k^\prime + \bm p^\prime - \bm k - \bm p)  \\
		\times
		\frac{1}{\sqrt{\mu(\bm k', \bm p') \mu(\bm k, \bm p)}} \, \hat f(\bm k'_\text{2cm}, \bm k_\text{2cm}),
		\label{<t>-t()}
	\end{multline}
	where $N(\bm p)$ denotes a single nucleon with momentum $\bf p$,
	$\bm k_\text{2cm}$ is the pion momentum in the pion-nucleon c.m. frame defined by the Lorentz transformation and $\mu(\bm k, \bm p) = {\omega(k)E_N(p)}/{W(\bm k, \bm p)}$ in terms of the nucleon energy $E_N(p)$ and the invariant mass $W(\bm k, \bm p)$ of the pion-nucleon system.
	The amplitude $\hat f$ is an operator in nucleon spin-isospin space analogous to the free-space pion-nucleon scattering amplitude.
	However, in the nuclear medium, it undergoes additional modifications due to the presence of numerous inelastic channels (see Section~IV.C in Ref.~\cite{Tsaran:2023qjx}).
	
	While the c.m.~momentum $\bm k_\text{2cm}$ and the phase factor $\mu$ depend explicitly on the nucleon momentum $\bm p$, we follow the standard \textit{optimized factorization approximation}~\cite{Landau:1975zb, Gurvitz:1986zza}, evaluating these quantities at the effective initial and final nucleon momenta
	\be
	\bm p_\text{eff} =  \frac{\bm q}{2} - \frac{\bm k^\prime + \bm k}{2A}, \quad\text{and}\quad
	\bm p_\text{eff}^\prime = - \frac{\bm q}{2} - \frac{\bm k^\prime + \bm k}{2A},
	\ee 
	where $\bm q = \bm k' - \bm k$.
	
	Under these assumptions, the first-order potential in momentum space takes the form~\cite{Gmitro:1985kj}
	\be
	V^{[1]}(\bm k', \bm k) = \mathscr{W}(\bm k', \bm k)
	\Tr \left[  \hat f\left(\bm k', \bm k\right) 
	\rho(\bm q) 
	\right],
	\label{V-1st-fin}
	\ee
	where the nuclear form factor $\rho(q)$ is normalized to $\Tr \rho(0) = A$, and 
	\be
	\mathscr{W}(\bm k', \bm k) =  \sqrt{ \frac{\mathscr{M}(k') \mathscr{M}(k)}{\mu(\bm k', \bm p_\text{eff}') \mu(\bm k, \bm p_\text{eff})}}
	\ee
	is a purely kinematic factor.
	Here and further, $\Tr$ represents the summation over all nucleon spin and isospin projections, which are omitted for simplicity.
	Similarly, the ``2cm" subscript in the arguments of the scattering amplitude $f$ is omitted but remains implicit.

	Considering the expectation value in Eq.~(\ref{V-all-def}) only over the nuclear ground state $\ket{\Psi_0}$, the structure of this contribution can be decomposed into two distinct terms
	\begin{multline}
		\langle  \Psi_0 | \hat f_2  \hat G \hat P_\emptyset \hat f_1 | \Psi_0 \rangle = - \langle \Psi_0 | \hat f_2 | \Psi_0 \rangle \hat G \langle \Psi_0 | \hat f_1 | \Psi_0 \rangle \\
		+ \langle \Psi_0 | \hat f_2 \hat G \hat f_1 | \Psi_0 \rangle . 
		\label{second-ord-term-general}
	\end{multline}
	Here, the first term involves the product of two expectation values of the pion-nucleon scattering amplitude.
	Each matrix element $\langle \Psi_0 | \hat f_2 | \Psi_0 \rangle$ is related to the first-order potential given in Eq.~(\ref{V-1st-fin}) (see Ref.~\cite{Tsaran:2023qjx} for details). 
	As a result, in momentum space, the part of the potential corresponding to the first term in Eq.~(\ref{second-ord-term-general}) takes the form
	\begin{multline}
		\tilde V^{[2]}(\bm k' ,\bm k) 
		= \frac{A-1}A  \int  \frac{\diff \bm k''}{2\pi^2}   \frac{1}{k_0^2 - {k''}^2+ i\varepsilon} \\
		\times
		V^{[1]}(\bm k' ,\bm k'') V^{[1]}(\bm k'' ,\bm k).
		\label{U2-1st-part}
	\end{multline}

	Unlike the first term, the second term in Eq.~(\ref{second-ord-term-general}) cannot be reduced to a product of one-body matrix elements 
	as it represents the expectation value of a two-body operator $\hat f_2 \hat G \hat f_1$.
	
	In momentum space, this term can be expressed as
	\begin{multline}
		\langle \pi(\bm k') \Psi_0 |\hat f_2 \hat G \hat f_1| \pi(\bm k) \Psi_0 \rangle  = 
		\frac1{ A(A-1)} \int \frac{\diff \bm k''}{(2\pi)^3}  G(k'') \\
		\times
		\Tr\left[ \hat
		f_2(\bm k^{\prime} ,\bm k^{\prime\prime})  
		\hat f_1(\bm k^{\prime\prime} ,\bm k) 
		\rho(\bm k^\prime - \bm k^{\prime\prime},  \bm k^{\prime\prime} - \bm k)
		\right],
		\label{U2-2nd-part}
	\end{multline}
	where $\rho(\bm q_1,  \bm q_2)$ is the two-body nucleon density matrix in momentum space, normalized to $\Tr \rho(0, 0) = A(A-1)$.
	In coordinate space, the latter is defined as
	\begin{multline}
		\rho(\bm r_1, \sigma_1, \tau_1, \bm r_2, \sigma_2, \tau_2 ; \bm r_1', {\sigma_1}', {\tau_1}', \bm r_2', {\sigma_2}', {\tau_2}')  \\
		= \langle \Psi_0 | 
		\hat \psi^\dag(\bm r_1, \sigma_1, \tau_1)\hat \psi^\dag(\bm r_2, \sigma_2, \tau_2)
		\\ \times
		\hat \psi(\bm r_1', {\sigma_1}', {\tau_1}')\hat \psi(\bm r_2', {\sigma_2}', {\tau_2}')
		| \Psi_0 \rangle,
		\label{two-body-density-matrix-(adag,a)}
	\end{multline}
	where $\hat \psi^\dag(\bm r, \sigma, \tau)$ and $ \hat \psi(\bm r, \sigma, \tau)$ are the creation and annihilation operators for a nucleon at position $\bm r$ with spin projection $\sigma$ and isospin projection $\tau$.
	General definitions and properties of the one- and two-body density matrices are considered in Appendix~\ref{sec:density-matrices-def}.
	Since we are considering target nuclei with zero angular momentum, we can trace over spin projections and define, for later convenience, the following spin-independent density matrix
	\begin{multline}
		\rho(\bm q_1, \bm q_2 ; \tau_1, \tau_2;  {\tau_1}', {\tau_2}') = 
		\sum_{\sigma_{1,2}}
		\int \! \diff \bm r_1 \! \diff \bm r_2 \, e^{- i (\bm q_1 \cdot \bm r_1 + \bm q_2 \cdot \bm r_2)}  
		\\ \times
		\rho(\bm r_1, \sigma_1, \tau_1, \bm r_2, \sigma_2, \tau_2; \bm r_1, \sigma_1, {\tau_1}', \bm r_2, \sigma_2, {\tau_2}'),
		\label{2b-density-mom-sp}
	\end{multline}

	The general form of the potential described in this subsection was employed in Ref.~\cite{Tsaran:2023qjx} to derive the scattering potential for spin-zero nuclei with equal numbers of protons and neutrons, achieving a good description of experimental data for $\pi^\pm$ scattering on ${}^{12}$C, ${}^{16}$O, ${}^{28}$Si, and ${}^{40}$Ca.
	For these nuclei, isospin invariance was assumed, enabling a simplified treatment of the spin-isospin summation in Eq.~(\ref{U2-2nd-part}) without the need to distinguish between proton and neutron distributions. This no longer holds for nuclei with non-zero isospin, such as $^{48}$Ca.
	
	\subsection{The non-zero isospin case}
	\label{sec:V-2nd}
	
	A comprehensive description of spin and isospin transitions during rescattering requires a detailed treatment of the spin-isospin structure of both nuclear structure functions and the pion-nucleon scattering amplitude.
	The latter can be expressed as the decomposition~\cite{Ericson:1988gk}
	\be
	\hat f = \hat f^{(0)} + \hat f^{(1)} \ \hat{\bm t} \cdot \hat{\pmb \tau} + 
	\left(\hat f^{(2)} +
	\hat f^{(3)} \ \hat{\bm t} \cdot \hat{\pmb \tau} \right) \hat{\pmb \sigma} \cdot \bm n,
	\label{t-isospin-struct}
	\ee
	where $\hat{\bm t}$ and $\hat{\pmb \tau}$ are the pion and nucleon isospin operators, $\hat{\pmb \sigma}$ is the nucleon spin operator, and $\bm n$ is the unit normal to the scattering plane.
	We adopt the convention that protons (neutrons) have positive (negative) isospin projection.
	
	Substituting the expansion given by Eq.~\eqref{t-isospin-struct} into Eq.~\eqref{V-1st-fin}, yields the first-order potential
	\begin{multline}
		V^{[1]}(\bm k', \bm k) = \mathscr{W}(\bm k', \bm k) 
		\left( 
		f^{(0)} (\bm k', \bm k) \left[\rho_n(q) + \rho_p(q) \right] 
		\right. \\ \left.
		- e_\pi f^{(1)} (\bm k', \bm k) \left[\rho_n(q) - \rho_p(q) \right]
		\right),
		\label{V(1)-fin}
	\end{multline}
	where $\rho_n(\bm q)$ and $\rho_p(\bm q)$ are the neutron and proton form factors, while $e_\pi$ is the pion charge.
	The form factors are normalized to the number of the corresponding nucleons, such that $\rho_p(q=0)=Z$ and $\rho_n(q=0)=N$.
	
	To simplify the cumbersome derivation of $V^{[2]}(\bm k' ,\bm k)$, we will break down the key steps rather than directly substituting Eq.~(\ref{t-isospin-struct}) into Eq.~(\ref{U2-2nd-part}) in momentum-space representation.
	The primary distinction from the isospin-zero case considered in Ref.~\cite{Tsaran:2023qjx} lies in the summation over isospin projections for nuclei with non-zero isospin. 
	To address this, we first examine the isospin structure involved in this summation. Specifically, we consider
	\begin{widetext}
		\begin{multline}
			\Tr_\tau \left[  \left( 
			\hat f^{(0)}_2 + \hat f^{(1)}_2 \ \hat{\bm t} \cdot \hat{\pmb \tau}_2 
			\right)  \left( 
			\hat f^{(0)}_1 + \hat f^{(1)}_1 \ \hat{\bm t} \cdot \hat{\pmb \tau}_1 
			\right) \rho \right] 
			= 
			2 \left( \hat f^{(0)}_2 \hat f^{(0)}_1 - \hat f^{(1)}_2 \hat f^{(1)}_1 \right )  \rho_{np} 
			\\
			+  \left(\hat f^{(0)}_2 + e_\pi \, \hat f^{(1)}_2 \right ) \left( \hat f^{(0)}_1 + e_\pi \, \hat f^{(1)}_1 \right )  \rho_{pp}  
			+ \left( \hat f^{(0)}_2 - e_\pi \, \hat f^{(1)}_2 \right ) \left( \hat f^{(0)}_1 - e_\pi \, \hat f^{(1)}_1 \right )  \rho_{nn} 
			+ 2 \hat f^{(1)}_2 \hat f^{(1)}_1 C_\text{(ex)}, 
			\label{Tr2-isospin}
		\end{multline}
	\end{widetext}
	where $\rho_{np},\rho_{pp}$, and $\rho_{nn}$ are the diagonal elements of  Eq.~\eqref{2b-density-mom-sp}  defined by setting  $\tau_1' =\tau_1$ and $\tau_2'=\tau_2$. 
	In this regard, we define $\rho_{\tau_1 \tau_2}(\bm q_1, \bm q_2) = \rho(\bm q_1, \bm q_2; \tau_1, \tau_2; \tau_1, \tau_2)$, where $\tau_1 \tau_2 = pp,nn,np$ specifies the isospin projections of the nucleon pair.
	The corresponding terms in Eq.~(\ref{Tr2-isospin}) describe pion scattering on a pair of nucleons while preserving the individual nucleon isospin.
	In contrast, the first term on the right-hand side of Eq.~(\ref{Tr2-isospin}) accounts for the intermediate charge-exchange process, which swaps the isospin projections in the $np$ pair.
	The associated density matrix is off-diagonal in isospin, and we refer to it as the exchange correlation function, defined as 
	\be
	C_\text{(ex)}(\bm q_1, \bm q_2)  =  -\rho(\bm q_1, \bm q_2; n, p; p, n),
	\ee
	with normalization $C_\text{(ex)}(0, 0) = \text{min}(N, Z)$.
	The factor of 2 in the second term of Eq.~(\ref{Tr2-isospin}) arises from the symmetry of this term under the exchange of $p$ and $n$.
	
	Although we are considering nuclei with zero spin, the intermediate spin-exchange dynamics results in a nontrivial outcome for the spin summation~\cite{Tsaran:2023qjx}.
	Due to spin invariance, the summation with the spin-dependent part of the amplitudes can be factorized, allowing us to consider separately
	\begin{align}
		\label{Tr2-spin}
		& \Tr_\sigma \left[ \left( 
		\hat f^{(0)}_2 + \hat f^{(2)}_2 \ \hat{\pmb \sigma}_2 \cdot \bm n_2
		\right)  \left( 
		\hat f^{(0)}_1 + \hat f^{(2)}_1 \ \hat{\pmb \sigma}_1 \cdot \bm n_1 
		\right)  \right] \nonumber  \\
		& \qquad\qquad\qquad\ \ \ 
		= 2 \left(\hat f^{(0)}_2 \hat f^{(0)}_1 + \hat f^{(2)}_2 \hat f^{(2)}_1 \, \bm n_1 \cdot \bm n_2 \right),  
	\end{align}
	where $\bm n_1 = \bm k_\text{2cm} \times \bm k_\text{2cm}''/|\bm k_\text{2cm} \times \bm k_\text{2cm}''|$ and $\bm n_2 = \bm k_{\text{2cm}'}'' \times \bm k_{\text{2cm}'}'/|\bm k_{\text{2cm}'}'' \times \bm k_{\text{2cm}'}'|$, with the subscript "2cm" ("$\text{2cm}'$") corresponding to the c.m. system of the pion and the first (second) nucleon.  
	
	Combining Eqs.~(\ref{Tr2-isospin}) and~(\ref{Tr2-spin}) to derive the full second-order part of the potential in momentum space, we obtain
	\begin{multline}
		V^{[2]}(\bm k' ,\bm k) 
		= \int \frac{\diff \bm k''}{2\pi^2}   \frac{\mathscr{W}(\bm k', \bm k'') \mathscr{W}(\bm k'', \bm k)  }{k_0^2 - {k''}^2+ i\varepsilon} 
		\\  \times
		\left(
		\varv_\text{(ex)} + \varv_{nn} + \varv_{pp} + 2 \varv_{np}
		\right) 
		-\frac1{A-1} \tilde V^{[2]}(\bm k' ,\bm k),
		\label{V(2)-fin}
	\end{multline}
	where the charge-exchange part of the potential is given by
	\begin{multline}
		\varv_\text{(ex)} = 2 \left[ f^{(1)}(\bm k' ,\bm k'') f^{(1)}(\bm k'' ,\bm k)
		\right. \\ \left.
		+ f^{(3)}(\bm k' ,\bm k'') f^{(3)}(\bm k'' ,\bm k) \, \bm n_1 \cdot \bm n_2
		\right]
		C_\text{(ex)}(\bm q_1,  \bm q_2),
		\label{v_ex}
	\end{multline}
	and the terms $\varv_{nn}$, $\varv_{pp}$, and $\varv_{np}$, which preserve individual nucleon isospin, are defined by
	\begin{widetext}
		\begin{multline}
			\varv_{\tau_1\tau_2} = 
			\left[\left(f^{(0)}(\bm k' ,\bm k'') + e_\pi \varepsilon_{\tau_1} f^{(1)}(\bm k' ,\bm k'') \right)
			\left(f^{(0)}(\bm k'' ,\bm k) + e_\pi \varepsilon_{\tau_2} f^{(1)}(\bm k'' ,\bm k)
			\right)
			\right. \\ \left.
			+ \left(f^{(2)}(\bm k' ,\bm k'') + e_\pi \varepsilon_{\tau_1} f^{(3)}(\bm k' ,\bm k'') 
			\right)
			\left(f^{(2)}(\bm k'' ,\bm k) + e_\pi \varepsilon_{\tau_2} f^{(3)}(\bm k'' ,\bm k) 
			\right) \bm n_1 \cdot \bm n_2
			\right] C_{\tau_1\tau_2}(\bm q_1,  \bm q_2),
			\label{v_tautau}
		\end{multline}
	\end{widetext}
	with $\varepsilon_p = 1$ and $\varepsilon_n = -1$.
	Here, we have introduced the correlation functions for a nucleon pair, as given by 
	\be
	C_{\tau_1\tau_2}(\bm q_1, \bm q_2)  = \rho_{\tau_1} (q_1) \rho_{\tau_2} (q_2) - \rho_{\tau_1\tau_2}(\bm q_1, \bm q_2),
	\label{C-all-def}
	\ee
	which are normalized to $C_{\tau_1\tau_2}(0,0) = \delta_{\tau_1 \tau_2} A_{\tau_1}$.
	We are using the spin-independent operator Eq.~\eqref{2b-density-mom-sp}, so the factor 2 in Eq.~\eqref{Tr2-spin} is already accounted for in the correlation functions.
	The last term in Eq.~(\ref{V(2)-fin}) accounts for the distinction between the processes described by $\varv_{(\text{ex})}$ and $\varv_{\tau_1 \tau_2}$, from the point of view of the Pauli exclusion principle.

	The resulting second-order potential for pion scattering on spin-zero nuclei, derived with explicit treatment of the intermediate spin-isospin dynamics, is given by Eqs.~(\ref{V(1)-fin}) and~(\ref{V(2)-fin})--(\ref{v_tautau}).
	For nuclei with zero isospin, assuming isospin invariance and the Slater determinant form for the nuclear ground-state wave function, the two-nucleon correlation functions satisfy $C_{pp}(\bm q_1, \bm q_2) = C_{nn}(\bm q_1, \bm q_2) = C_\text{(ex)}(\bm q_1, \bm q_2)$ and $C_{np} = 0$.
	Under this condition, Eqs.~(\ref{V(2)-fin})--(\ref{v_tautau}) simplify to Eq.~(52) from Ref.~\cite{Tsaran:2023qjx}.
	Note that in Eq.~(\ref{V(2)-fin}), $\varv_{np}$ vanishes under the Slater determinant assumption employed in the following, as $C_{np}$ is zero in this case (see Section~\ref{sec:V-2nd}).

	The calculation of the second-order potential requires two types of inputs: the scattering amplitudes $f^{(i)}$ and the nuclear structure functions $\rho_{\tau}(q)$ and $C_{\tau_1 \tau_2, (\text{ex})}(\bm{q}_1, \bm{q}_2)$.
	The scattering amplitudes on a single nucleon are universal across all nuclei but include free parameters that account for their modification within the nuclear medium. 
	In our approach, the free-space amplitudes are taken from the phase shift analysis (WI08) by the SAID Collaboration~\cite{Workman:2012hx}, while their in-medium modifications are characterized by three real energy-independent parameters.
	The three parameters consist of the real and imaginary parts of the $\Delta(1232)$ self-energy, which modify the resonant $P_{33}$ channel, and the slope parameter of the imaginary part of the $s$-wave isoscalar amplitude.
	These free parameters were determined in Ref.~\cite{Tsaran:2023qjx} by fitting $\pi^\pm$-${}^{12}$C scattering data in the pion laboratory kinetic energy range of 80--\SI{180}{MeV}, which shows a strong sensitivity to the $\Delta$-resonance properties.
	These parameters are adopted unchanged in the present analysis.
	
	In contrast to the single-nucleon scattering amplitudes, the nuclear structure functions are specific to each nucleus.
	Rather than being fitted to pion-nucleus scattering data, they are either extracted from experiments or determined through theoretical predictions.
	In Ref.~\cite{Tsaran:2023qjx},  under the assumption of isospin invariance,  nucleon form factors were determined from the measured charge form factors, while two-body correlation functions were derived using a non-interacting harmonic oscillator (HO) shell model.

	Here, as follows from Eq.~(\ref{V(1)-fin}), constructing the potential for nuclei with non-zero isospin requires separate knowledge of both point-proton and point-neutron form factors.
	In particular, the distributions of neutrons and protons inside the nucleus differ. While the proton density can still be extracted from experimental charge densities, a precise, model-independent determination of neutron densities from experiment is challenging or accompanied by poorly controlled uncertainties~\cite{Piekarewicz:2005iu}.
	As a result, theoretical predictions are required not only for obtaining the correlation functions but also for the point-proton and point-neutron form factors. In this work, we take this input from modern nuclear structure theory.
	
	\section{Nuclear structure}
	\label{sec: nuclear structure}
	
	The foundation of \textit{ab initio} nuclear structure calculations~\cite{papenbrock2024,Hergert2020,Ekstrom2023} lies in the choice of the model for the interaction among nucleons and of an appropriate many-body method.
	Regarding the many-body technique, several different methods can be used~\cite{Hergert2020}. Here, we work with coupled-cluster (CC) theory~\cite{Hagen2014Review,ShavittBartlett,Hagen_2016_rev} and approximations thereof. CC theory stands out for the excellent accuracy at a moderate computational cost, allowing access to medium-mass or even heavy nuclei~\cite{PbAbInitio}, as well as electroweak observables~\cite{Bacca2013,Hagen2016,Payne:2019wvy,Sobczyk2021,Sobczyk:2023mey}.
	In terms of interactions, Hamiltonians obtained from chiral effective field theory ($\chi$EFT)~\cite{Epelbaum2009,machleidt2011,MACHLEIDT2024104117,Epelbaum2024}, including nucleon-nucleon and three-nucleon contributions, are employed.
	$\chi$EFT provides a systematic approach to derive nuclear Hamiltonians consistent with the symmetries of the underlying theory of QCD.
	The guiding principle is a low-momentum expansion of the interactions in powers $(Q/\Lambda_{\chi})^\nu$, where $Q$ is a typical low-momentum scale characterizing nuclear physics and $\Lambda_\chi \approx 1\,\rm{GeV}$ is the chiral symmetry breaking scale within QCD.
	~Chiral forces depend on a set of parameters, called low-energy constants (LECs), that are constrained by experimental data or theoretical results.
	
	In particular, we will use two interactions at the next-to-next-to-leading order in the $\chi$EFT expansion (i.e., including terms up to $\nu=3$), namely $\rm{NNLO_{sat}}(450)$~\cite{NNLOsat} and the $\Delta$-full model $\Delta \rm{NNLO_{GO}}(394)$~\cite{DeltaGo2020}, with the associated momentum cutoff (in MeV/c) shown in parenthesis. For $\rm{NNLO_{sat}}(450)$ the LECs were adjusted to reproduce nucleon-nucleon (NN) scattering data, as well as energies and radii of selected light and medium-mass nuclei up to $^{24}\rm{O}$, while $\Delta \rm{NNLO_{GO}}(394)$ has been fitted to properties of both light nuclei and nuclear matter in addition to NN phase shifts.
	Both models are able to accurately describe bulk observables (such as charge radii and densities) in the calcium region relevant for this work, see e.g.~\cite{Payne:2019wvy,PbAbInitio,DeltaGo2020,Soma2020Chiral}.
	By comparing the results obtained with these two different chiral forces, we can appreciate the effect of including explicit delta isobars in the chiral expansion, as well as of varying the cutoff and the optimization procedure employed for the LECs. 
	
	To discuss the many-body approach, we start from Hartree-Fock (HF) theory, which provides the simplest approximation to the g.s. wave function, assumed to be a single Slater determinant 
	\begin{align}
		\ket{\Phi_0} =
		\mathcal{A} \{ \ket{p_1} \otimes ... \otimes \ket{p_A} \},
	\end{align}
	where $\mathcal{A}$ is the antisymmetrization operator and the single-particle (s.p.) orbitals $\ket{p_j}$ are determined self-consistently.
	In the following, we will focus on closed-shell nuclei with g.s. angular momentum and parity $J^{\pi} = 0^{+}$.
	In this case, a spherical HF solution provides a reasonable starting point for a more advanced many-body calculation~\cite{Hagen2014Review}.
	
	In CC theory~\cite{Hagen2014Review,Hagen_2016_rev,ShavittBartlett} the g.s.~wave function  $\ket{\Psi_0}$ of a given nucleus is determined from the following exponential parametrization of the many-body g.s.~,
	\begin{align}
		\label{eq: cc gs ansatz}
		\ket{\Psi_0} = e^{ \hat{T} } \ket{ \Phi_0 }.
	\end{align}
	Dynamical correlations are built on top of the reference state $\ket{ \Phi_0 }$ by the action of $e^{ \hat{T} }$, where the cluster operator $\hat{T}$ is expanded as a combination of $n$-particle-$n$-hole ($n$p-$n$h) excitation operators.
	We employ a common truncation of the cluster operator, known as coupled-cluster at the singles and doubles level (CCSD), including terms in $\hat T$ up to $2p$-$2h$ excitations,
	\begin{align}
		\hat{T} = \sum_{ai} t^{a}_{i} \hat{c}_a^{\dagger} \hat{c}_i +
		\frac{1}{4} \sum_{abij} t^{ab}_{ij} \hat{c}_a^{\dagger} \hat{c}_b^{\dagger} \hat{c}_j \hat{c}_i + ...,
	\end{align}
	where $\hat{c}$ ($\hat{c}^{\dagger}$) are the annihilation (creation) operators and indices $i,j$ and $a,b$ denote single-particle states that are occupied (holes) and unoccupied (particles) in the reference state $\ket{ \Phi_0 }$, respectively. 
	The amplitudes $t^{a}_{i}$, $t^{ab}_{ij}$ etc. are determined by solving the CC equations. 
	We also investigate the effect of including leading $3p$-$3h$ contributions in the CCSDT-1 scheme~\cite{ShavittBartlett,Miorelli2018}.
	
	The HF state is built from a model space consisting of 15 major HO shells with frequency $\hbar\omega=16\,\rm{MeV}$ (22 MeV) for $\Delta \rm{NNLO_{GO}}(394)$ ($\rm{NNLO_{sat}}$]).
	These frequencies are optimal for describing charge radii and densities with these potentials (see~\cite{Hagen2016,DeltaGo2020}).
	In the following, Greek indices denote states in the s.p.~spherical HO basis, $\alpha \equiv \{n; l j m \tau \}$.%
	\footnote{
		Note that the subscript $n$ in the correlation functions $C_{nn, np}$ denotes neutrons, whereas $n$ in the radial wave function $R_{nl}$ and in the one-body density matrix $\rho_{j_\alpha l_\alpha \tau_\alpha }^{(n_\alpha n_\beta)}$ refers to the principal quantum number of a s.p.~state.
	}%
	Here, $n$ is the principal quantum number, $l$ and $j$ refer to the orbital and total angular momentum, respectively, $m$ is the angular momentum projection, and $\tau$ is the isospin projection of a nucleon.
	The HO s.p. wave functions in real space are given by~\cite{Vorabbi:2023mml,Scalesi2024}
	\begin{align}
		\label{eq: sp orbitals}
		\phi_{\alpha}(x) = \bra{x}\ket{\alpha} =  R_{nl}(r) \mathcal{Y}_{ljm}(\hat{\bm r},\sigma)\,,
	\end{align}
	where $x = (\bm r, \sigma)$ encompasses the (three-dimensional) position ($\bm r$) and spin projection ($\sigma$) degrees of freedom. 
	$R_{nl}(r)$ and $\mathcal{Y}_{ljm}(\hat{\bm r},\sigma)$ are the radial and angular parts of the wave function, respectively.

	\subsection{One-body densitites}
	From the CC wave function, expectation values of observables can be determined as described in~\cite{Hagen2014Review,Stanton1993}.
	For our purposes, we focus on the one-body density matrix, whose matrix elements in the HO basis are given by
	\begin{align}
		\rho^{\alpha}_{\beta} =
		\mel{\Psi_0}{ \hat{c}^{\dagger}_{\alpha } \hat{c}_{\beta} }{ \Psi_0}.
	\end{align}
	In spherical nuclei, $\rho^{\alpha}_{\beta}$ is block-diagonal for any choice of spherical s.p.~states, since conservation laws of angular momentum, parity $\pi = (-1)^{l}$, and isospin projection enforce~\cite{Malaguti,Soma2011}
	\begin{align}
		\label{eq: 1b density symmetry}
		\rho^{\alpha}_{\beta} = \delta_{j_\alpha j_\beta} \delta_{m_{\alpha} m_{\beta} }
		\delta_{l_\alpha l_\beta} 
		\delta_{\tau_{\alpha} \tau_{\beta} } 
		\rho_{j_\alpha l_\alpha \tau_\alpha }^{(n_\alpha n_\beta)}.
	\end{align}
	That is, the one-body density matrix mixes up states of the spherical s.p. basis that differ by the principal quantum number only. 
	The corresponding real-space point-neutron and point-proton densities $\rho_\tau(\bm r, \bm r^\prime)$ are given by~\cite{Malaguti,Vorabbi:2023mml}
	\begin{align}
		\label{eq: 1b dens mat}
		\rho_{\tau}(\bm r, \bm r^{\prime} ) &= 
		\sum_{\sigma \sigma^{\prime} } \sum_{\alpha\beta} 
		\phi_{\alpha}(x)^{\dagger} \rho_{\beta}^{\alpha} 
		\phi_{\beta}(x^{\prime}) \delta_{\tau_\alpha \tau} \delta_{\tau_\beta \tau} 
		\\ &= \sum_{jl n_1 n_2 }
		\frac{2j+1}{4\pi} \rho_{jl\tau}^{(n_1 n_2)}
		R_{n_1 l}(r) R_{n_2 l}(r^\prime) P_l( \hat{\bm r} \cdot \hat{\bm r}^{\prime} ),
		\nonumber
	\end{align}
	where $P_l$ are the Legendre polynomials of order $l$.
	The spherical number densities $\rho_\tau(r)$ (with $r=\abs{\bm r}$) are readily obtained by setting $\bm r = \bm r^{\prime} $ and $P_l =1$ in Eq.~\eqref{eq: 1b dens mat}.
	The Fourier transform of $\rho_\tau(r)$ defines the form factor $\rho_{\tau}(q)$
	\begin{align}
		\label{eq: def form factor}
		\rho_{\tau}(q) =
		\int \diff \bm r \, e^{i \bm q \cdot \bm r } \rho_{\tau}(\bm r ) =
		4\pi \int_{0}^{+\infty} \diff r r^2 j_{0}(qr) \rho_{\tau}(r),
	\end{align}
	with $j_{0}(qr) = \sin(qr)/(qr)$.
	
	We obtain intrinsic one-body densities by following the prescription of Refs.~\cite{Hagen2009,Hagen2010,Hagen2016}.
	The CC wave function factorizes approximately into an intrinsic part times a Gaussian c.m.~wave function. Thus, a deconvolution with respect to the Gaussian wave function allows finding the intrinsic one-body density $\rho_{p,n}^{\rm in}(q)$ from $\rho_{\tau}(q)$ in the laboratory frame.
	Finally, the elastic charge form factor $\rho_\text{ch}(q)$ is found by folding $\rho_{p,n}^{\rm in}(q)$ with the individual proton and (small) neutron electric form factors,
	\begin{align}
		\rho_\text{ch}(q) = G_{p}(q)  \rho_{p}^{\rm in}(q)  + G_{n}(q)  \rho_{n}^{\rm in}(q) .
	\end{align}
	As in Ref.~\cite{Sobczyk2020}, we use the Kelly's parametrizations~\cite{Kelly2004} for $G_{p,n}$.
	The charge density can be determined through the inverse transform,
	\begin{align}
		\label{eq: inverse Fourier rho ff}
		\rho_\text{ch}(r) = 
		\frac{1}{2\pi^2}
		\int \diff q \, q^2 j_{0}(qr) \rho_\text{ch}(q).
	\end{align}

	\begin{figure}[t!]
		\centering
		\includegraphics[width=\columnwidth]{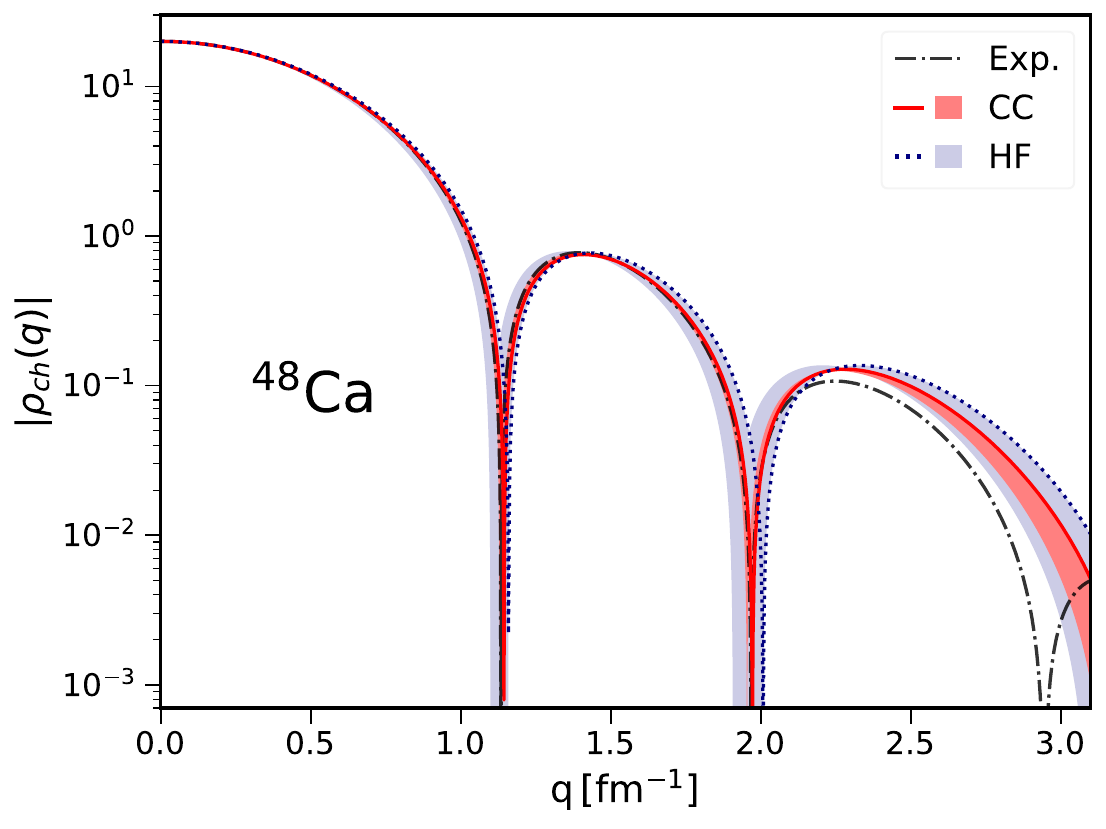}
		\caption{$^{48}\rm{Ca}$ charge form factors.
			Calculations performed with the CC method at the CCSDT-1 level and in HF theory using the $\Delta \rm{NNLO_{GO}(394)}$ interaction are shown as a full line and a dotted line, respectively.
			Shaded bands indicate the results obtained with the $\rm{NNLO_{sat}}(450)$ potential.
			Experimental data from Ref.~\cite{DeJager:1987qc} are also reported for comparison.
		}
		\label{fig: Form factors Ca48}
	\end{figure}
	Fig.~\ref{fig: Form factors Ca48} shows our results for the $^{48}\rm{Ca}$ charge form factor $\abs{ \rho_\text{ch}(q) }$ as a function of the momentum $q$ in comparison to the experimental data (dash-dotted line) from Ref.~\cite{DeJager:1987qc}.
	Computations performed with CC theory at the CCSDT-1 level with the $\Delta \rm{NNLO_{GO}(394)}$ interaction are shown as a full line.
	HF predictions with the same potential are reported as a dotted line.
	In both cases, the error bands account for the difference between the $\Delta \rm{NNLO_{GO}(394)}$ and $\rm{NNLO_{sat}}(450)$ interactions.
	We can appreciate that the agreement between CC predictions and the experiment is excellent up to the second diffraction minimum, and the interaction dependence remains mild up to almost 2.5 $\rm{fm}^{-1}$, at the edge of the region of validity of chiral interactions, as marked by their regularization cutoff~\cite{Payne:2019wvy}.
	Also, we notice that the data are described reasonably well already within the HF approximation, especially at low momentum.

	\subsection{The two-nucleon correlation functions}
	As demonstrated in Sect.~\ref {sec:multiple-scattering}, the nucleon-nucleon correlation functions and, correspondingly, the two-body density matrix play a key role in describing the second-order pion-nucleus optical rescattering. 
	In this work, we determine the two-body densities for ${}^{48}$Ca within the simplifying assumption of an HF wave function.
	This approach offers a good balance between accuracy and simplicity. 
	As shown in Fig.~\ref{fig: Form factors Ca48}, the HF framework reasonably describes one-body form factors. 
	Therefore, we expect the error related to using the HF state instead of the CC one in the two-body densities to be relatively small.
	At the same time, computations at the HF level are significantly less demanding than a CC calculation of two-body density matrices (see, e.g.,~\cite{Stanton1993, ShavittBartlett, Gauss2001}).

	Within the HF approach, the two-body density operators defined in Eq.~\eqref{two-body-density-matrix-(adag,a)} are evaluated on the reference Slater determinant state $\ket{\Phi_0}$. 
	As a consequence of Wick's theorem, the two-body density matrix $\rho(\bm q_1, \bm q_2; \tau_1, \tau_2; \tau_1, \tau_2)$ in this case can be expressed entirely in terms of the one-body density matrix, as shown in Appendix~\ref{app:Slater}.
	As a result, the correlation functions that enter the second-order potential, namely $C_{(\text{ex})}$ and $C_{\tau_1 \tau_2}$, can all be expressed in terms of a single function $D_{\tau_1 \tau_2}$, which is constructed directly from the one--body densities in coordinate space as
	\begin{align}
		D_{\tau_1 \tau_2}(\bm r_1, \bm r_2) = 
		\frac12
		\rho_{\tau_1}(\bm r_1, \bm r_2) 
		\rho_{\tau_2}(\bm r_2, \bm r_1).
		\label{eq: D func correlation}
	\end{align}
	Explicitly, the relationships among these quantities are given by
	\begin{subequations}
		\begin{align}
			\label{eq: C and D corr HF}
			& C_{(\text{ex})}(\bm q_1, \bm q_2) = D_{np}(\bm q_1, \bm q_2),\\
			& C_{pp,nn}(\bm q_1, \bm q_2) = D_{pp, nn}(\bm q_1, \bm q_2), \\
			& C_{np}(\bm q_1, \bm q_2) = 0,
		\end{align}
		\label{C(D)}
	\end{subequations}
	Upon applying a double Fourier transform analogous to Eq.~(\ref{2b-density-mom-sp}), the partial-wave decomposition of $D_{\tau_1 \tau_2}$ in momentum space takes the form
	\begin{align}
		D_{\tau_1 \tau_2}(\bm q_1, \bm q_2) = \sum_{L} (2L+1) D^{(L)}_{\tau_1 \tau_2}(q_1,q_2) P_L(x)  ,
		\label{eq: D func partial waves}
	\end{align}
	where $x = \bm q_1 \cdot \bm q_2 / (q_1 q_2)$.
	Here, the partial-wave components are given by
	\begin{multline}
		D^{(L)}_{\tau_1 \tau_2} (q_1,q_2) =
		\frac{(-1)^L}{2} \sum_{\substack{jl n_1 n_2 \\ j^\prime l^\prime n_1^\prime n_2^\prime}} 
		(2j+1) (2j^\prime+1) \begin{pmatrix}
			l & l^{\prime} & L \\ 0 & 0 & 0
		\end{pmatrix}^2  \\
		\times \chi_l \chi_{l'}
		Q^{(L)}_{n_1 l n_1^\prime l^\prime}(q_1)
		Q^{(L)}_{n_2 l n_2^\prime l^\prime}(q_2)
		\,
		\rho_{jl \tau_1}^{(n_1 n_2)}
		\rho_{j^\prime l^\prime \tau_2}^{ (n_1^\prime n_2^\prime) }, 
	\end{multline}
	where $\begin{pmatrix} l & l^{\prime} & L \\ 0 & 0 & 0 \end{pmatrix}$ is the Wigner $3j$ symbol.
	The factor $\chi_l$  accounts for the outer shell occupation:  it equals 1 for fully filled shells and $(2l + 1)/(l+1)$ for the outermost shell, when only its subshell%
	\footnote{A "subshell" refers to a group of orbitals sharing the same $n$, $l$, and $j$ quantum numbers, e.g., the $1p$ shell consists of the $1p_{3/2}$ and $1p_{1/2}$ subshells.} 
	is filled.
	This correction arises from differences in occupation probabilities between the $ls$ and $jj$ coupling schemes.
	The latter is used in the exchange sum for the outer subshell and ensures proper normalization~\cite{Jackson:1970zz} (see Ref.~\cite{murugesu1971optical} for a detailed derivation).

	Finally, the radial integrals $Q^{(L)}$ are defined as
	\begin{align}
		Q^{(L)}_{nl n^\prime l^\prime}(q) =
		\int_{0}^{+\infty} r^2 \! \diff r \, j_L(qr) R_{nl}(r)  R_{n^\prime l^\prime}(r) ,
	\end{align}
	with $j_L(x)$ being spherical Bessel functions. 
	The special case of a HO shell model, discussed in Ref.~\cite{Tsaran:2023qjx}, is recovered by setting $\rho^{\alpha}_\beta = \rho^{\alpha} \delta_{\alpha \beta}$ in Eq.~\eqref{eq: 1b density symmetry} and filling only the lowest $A$ HO states, i.e., $\rho^{\alpha}=1$ for occupied states and 0 otherwise.
	
	\begin{figure*}[!t]
		\includegraphics[width=0.5\textwidth]{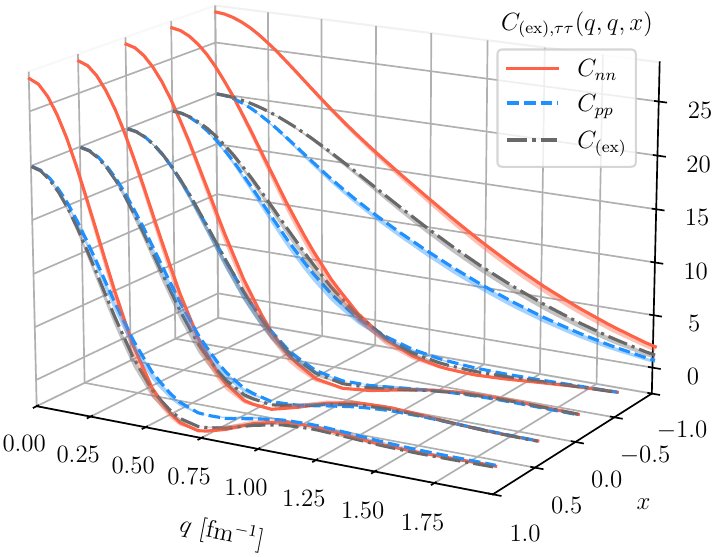}
		\hfill
		\includegraphics[width=0.49\textwidth]{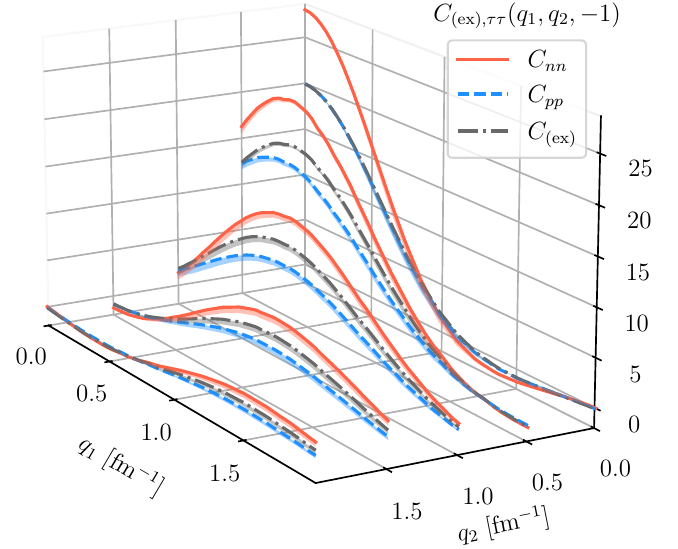}
		\includegraphics[width=0.5\textwidth]{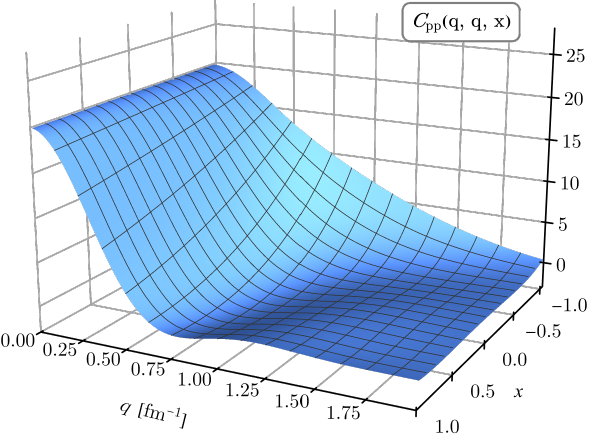}
		\hfill
		\includegraphics[width=0.49\textwidth]{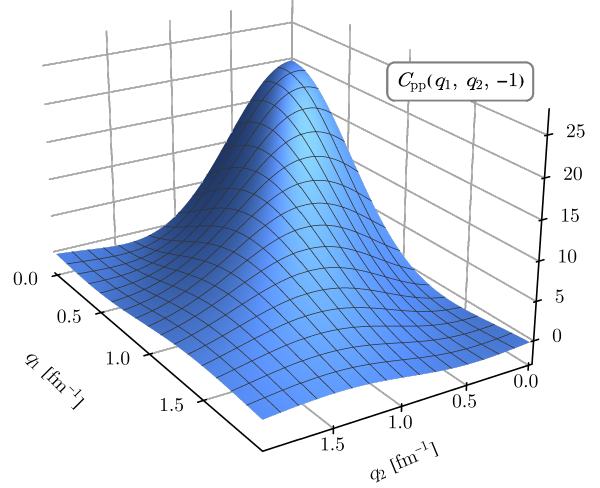}
		\caption{ The HF correlation functions $C_{(\text{ex}), \tau\tau}(\bm q_1, \bm q_2)$ for ${}^{48}$Ca are shown for two momentum configurations: 
			$|\bm q_1| = |\bm q_2| = q$ and $x = \bm q_1 \cdot \bm q_2 / (q_1 q_2)$ (left panels) and $|\bm q_1| = q_1$, $|\bm q_2| = q_2$ for $x=-1$ (right panels).
			The upper panels display the correlation functions with solid red, dashed blue, and dash-dotted gray lines corresponding to calculations using the $\Delta \rm{NNLO_{GO}}(394)$ interaction.
			The associated error bars indicate results obtained with the $\rm{NNLO_{sat}}(450)$ interaction.
			The lower panels present the same $\Delta \rm{NNLO_{GO}}(394)$ results for $C_{pp}$ as surface plots.
		}
		\label{fig:C_all_48Ca}
	\end{figure*}

	The two-nucleon correlation functions for ${}^{48}$Ca, obtained using the described HF approach, are shown in Fig.~\ref{fig:C_all_48Ca}. 
	As seen in the upper panels, the functions $C_{(\text{ex})}$, $C_{nn}$, and $C_{pp}$ exhibit similar overall shapes but are clearly distinguishable, showing a strong dependence on both momentum variables and the angle between them.
	All three correlation functions derived from the $\Delta \text{NNLO}_{\text{GO}}(394)$ and $\text{NNLO}_{\text{sat}}(450)$ interactions are in reasonable agreement.
	The associated error bands, reflecting uncertainties due to the choice of interaction, remain small relative to the overall variation of the correlation functions across the considered kinematic range.
	For enhanced visualization, the lower panels of Fig.~\ref{fig:C_all_48Ca} present surface plots for $C_{pp}$. 
	As follows from the upper panels, $C_{nn}$ and $C_{(\text{ex})}$ share similar qualitative features.
	
	The HF approach naturally extends and refines the non-interacting HO shell model used in Ref.~\cite{Tsaran:2023qjx} for nuclei with zero isospin.
	As a result, for nuclei where both approaches are applicable, the HF and non-interacting HO shell model ground-state wave functions yield closely matching correlation functions within the dominant momentum transfer range of $\SI{1.5}{fm^{-1}}$, with differences falling within the HF error bars.
	In particular, both approaches produce nearly identical results for ${}^{40}$Ca, for which pion scattering was successfully described in Ref.~\cite{Tsaran:2023qjx} using model parameters fitted to ${}^{12}$C pion-nucleus scattering data.
	This suggests that the same pion-nucleon scattering amplitudes and parameters can also be adopted for the potential developed in Sect.~\ref{sec:V-2nd}, when used in combination with the HF correlation functions.
	More details considering the case of ${}^{40}$Ca, are provided in Appendix~\ref{sec:40Ca}.

	\section{Results and discussion}
	\label{sec:fits-results}
	
	In Section~\ref{sec:multiple-scattering}, we have established the pion-nucleus potential for spin-zero nuclei with non-zero isospin, such as ${}^{48}$Ca.
	Using the nuclear-structure inputs for ${}^{48}$Ca as discussed in Sect.~\ref{sec: nuclear structure}, we are now in a position to test the predictive power of our model.
	We compare our theoretical predictions for differential elastic cross sections to data measured within the $\Delta$-resonance energy region at Schweizerisches Institut f\"ur Nuklearforschung (SIN)~\cite{Gretillat:1981bq} and  Los Alamos Meson Physics Facility (LAMPF)~\cite{Boyer:1984bs}.
	We note that the results shown in this section are based on the model parameters determined in Ref.~\cite{Tsaran:2023qjx} without any additional adjustments.

	While in Sect.~\ref{sec:multiple-scattering} we focused solely on the nuclear potential generating the hadronic amplitude, the interaction of charged pions with the nucleus is significantly influenced by electromagnetic forces.
	We incorporate the short- and long-range Coulomb effects in the full pion-nucleus scattering amplitude as described in Ref.~\cite{Tsaran:2023qjx} (see Sec.~V~A and Appendix~A therein).
	In the calculation, we account for the charge distribution of ${}^{48}$Ca adopting the model-independent Fourier-Bessel series, with expansion coefficients taken from Ref.~\cite{DeJager:1987qc}.
	
	\begin{figure*}[!t]
		\includegraphics[width=0.495\linewidth]{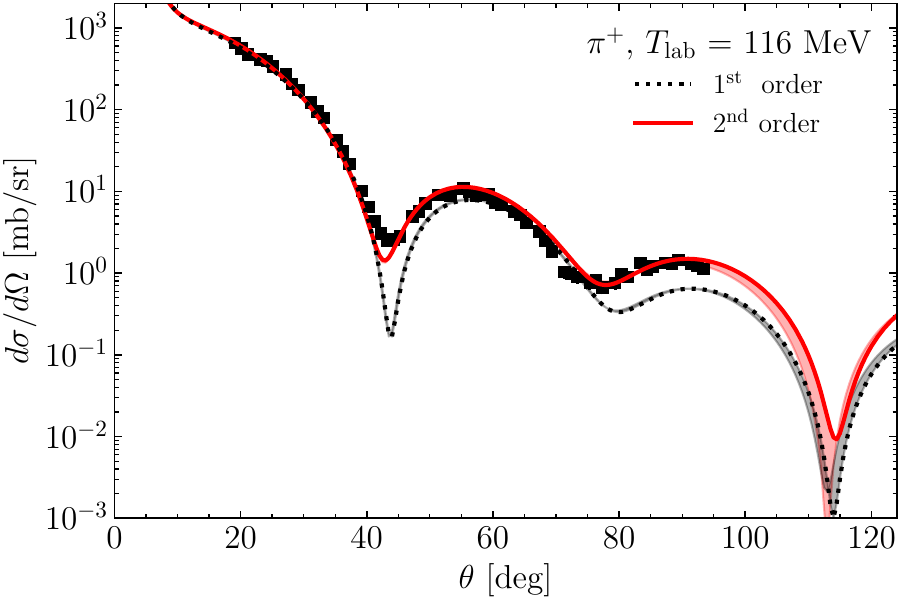}
		\hfill
		\includegraphics[width=0.495\linewidth]{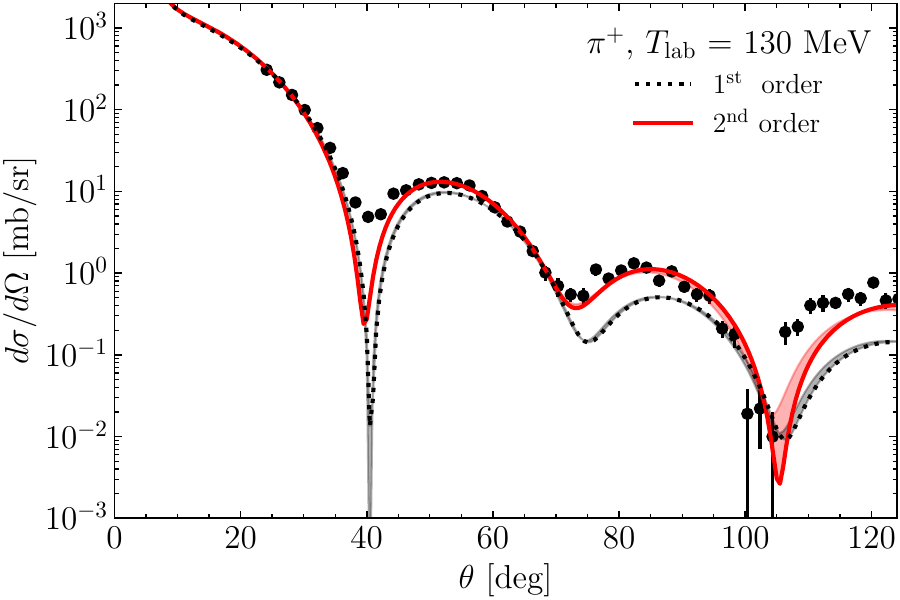}
		\includegraphics[width=0.495\linewidth]{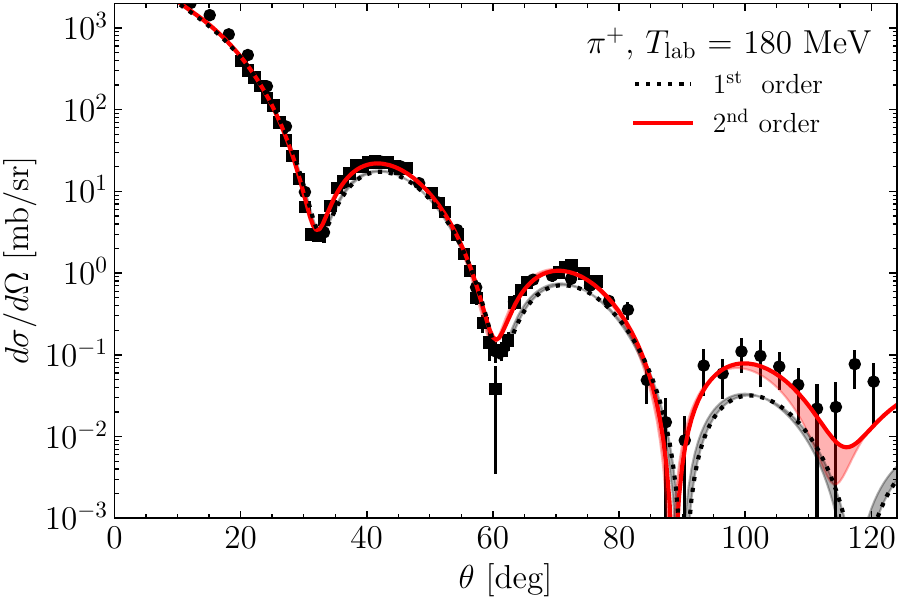}
		\hfill
		\includegraphics[width=0.495\linewidth]{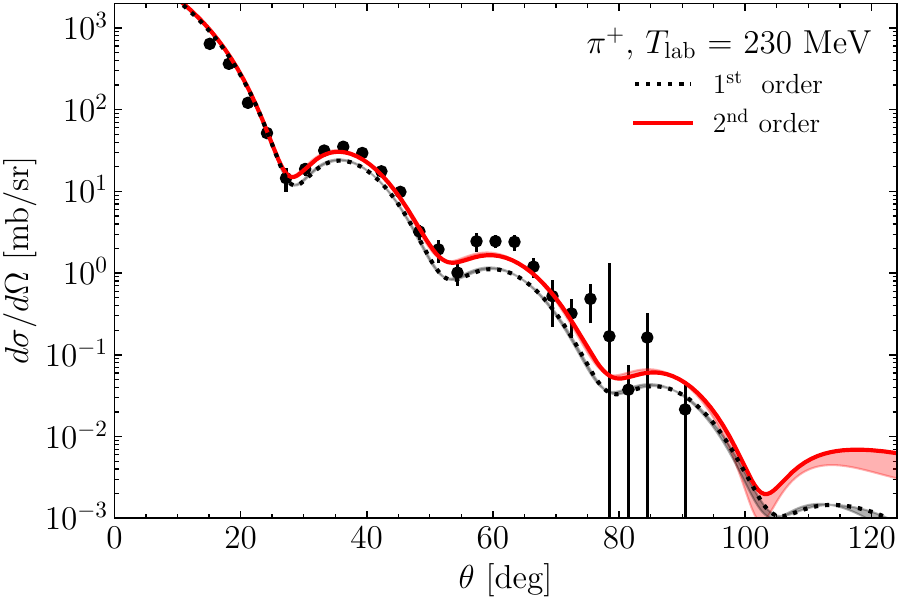}
		\caption{
			Predictions for differential cross sections for $\pi^+$-${}^{48}$Ca scattering, obtained with the first-order (short-dashed black curves) and second-order potentials (solid red curves), compared with data of Refs.~\cite{Boyer:1984bs}  ({\tiny  $\blacksquare$} at 116, $\SI{180}{MeV}$) and~\cite{Gretillat:1981bq} ({\large $\bullet$} at 130, 180, $\SI{230}{MeV}$).
			The theoretical curves are based on nucleon form factors derived from the CC theory and, for the second-order case, HF correlation functions, all derived using the $\Delta \rm{NNLO_{GO}}(394)$ interaction. 
			Error bars reflect the variation in results obtained when nuclear structure inputs from the $\rm{NNLO_{sat}}(450)$ interaction are used instead.
		}
		\label{plt:48CaP_1st}
	\end{figure*}
	
	\begin{figure*}[!t]
		\includegraphics[width=0.495\linewidth]{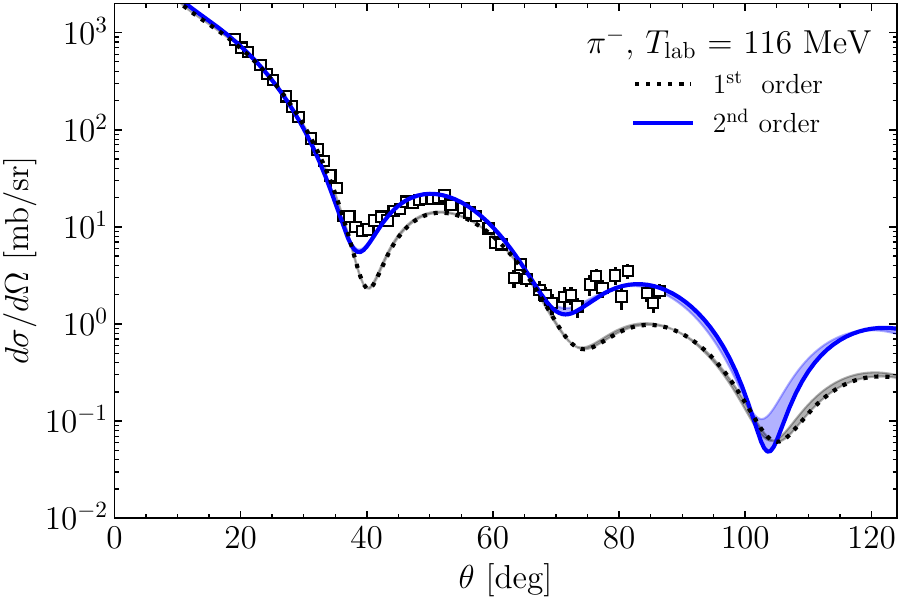}
		\hfill
		\includegraphics[width=0.495\linewidth]{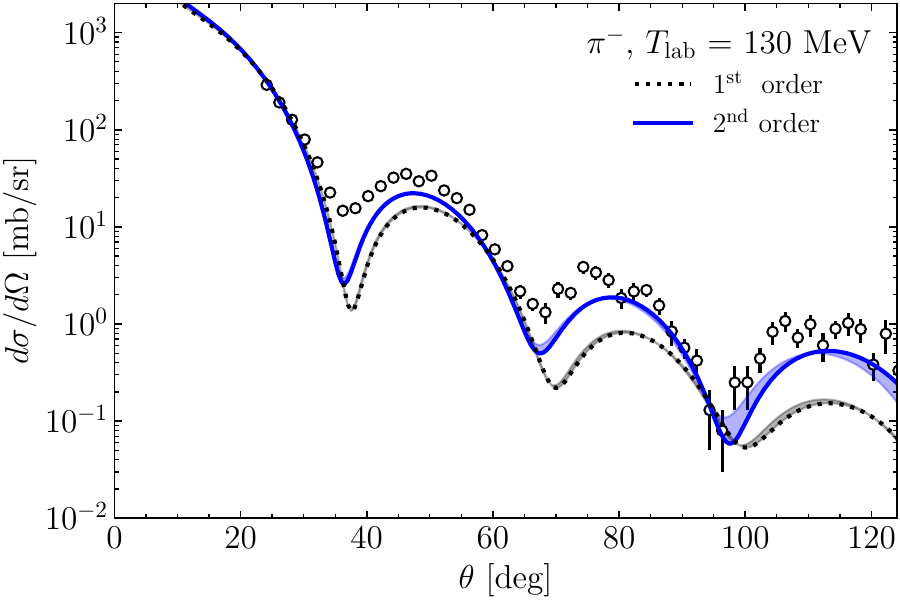}
		\includegraphics[width=0.495\linewidth]{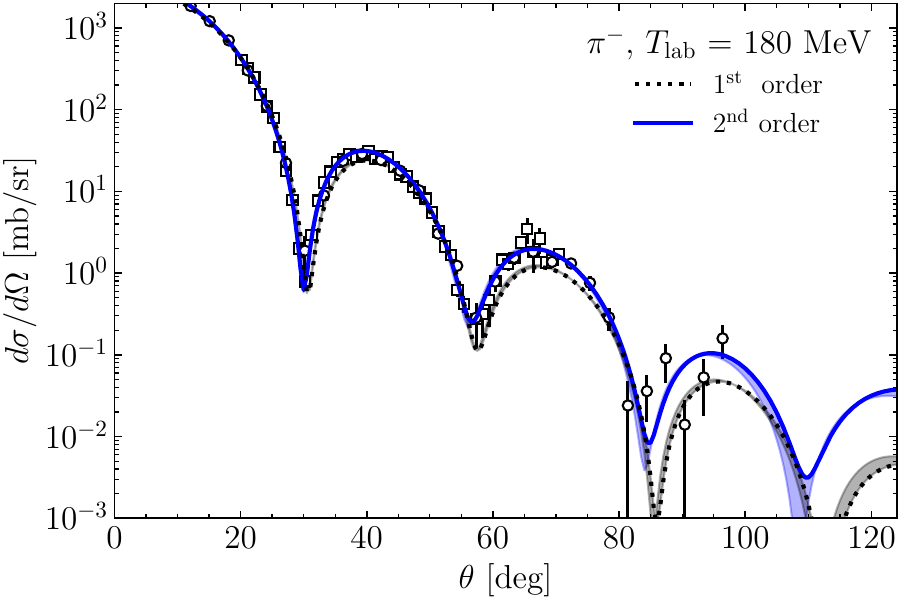}
		\hfill
		\includegraphics[width=0.495\linewidth]{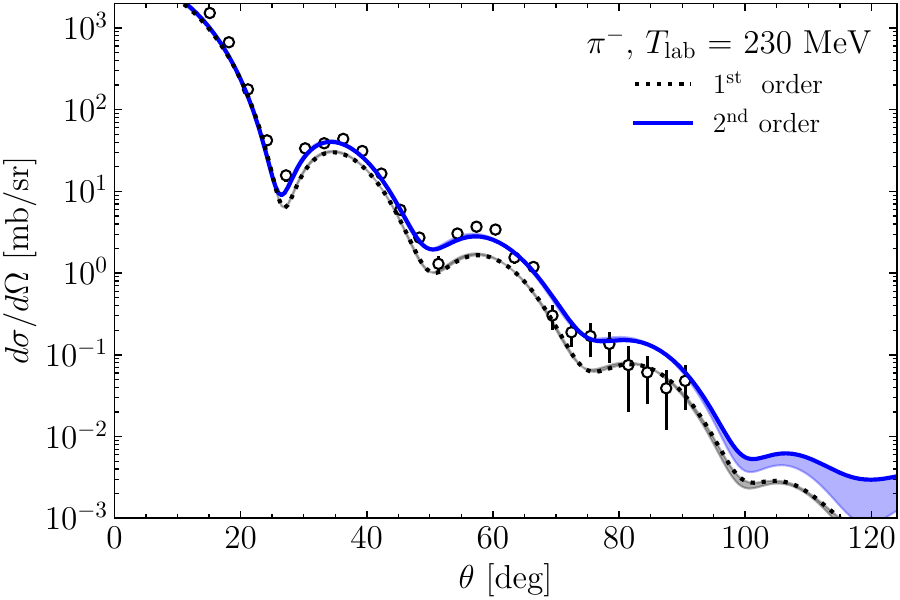}
		\caption{
			Predictions for differential cross sections for $\pi^-$-${}^{48}$Ca scattering, obtained with the first-order (short-dashed black curves) and second-order potentials (solid blue curves), compared with data of Refs.~\cite{Boyer:1984bs}  ({\tiny  $\square$} at 116, $\SI{180}{MeV}$) and~\cite{Gretillat:1981bq} ({\large $\circ$} at 130, 180, $\SI{230}{MeV}$).
			The curves and bands have the same meaning as in Fig.~\ref{plt:48CaP_1st}.
		}
		\label{plt:48CaM_1st}
	\end{figure*}
	
	\begin{figure*}[!t]
		\includegraphics[width=0.495\textwidth]{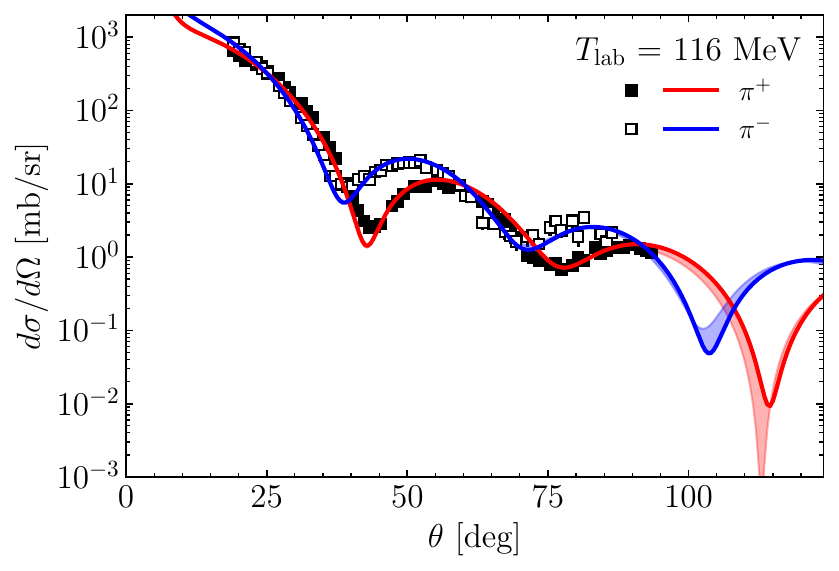}
		\hfill
		\includegraphics[width=0.495\textwidth]{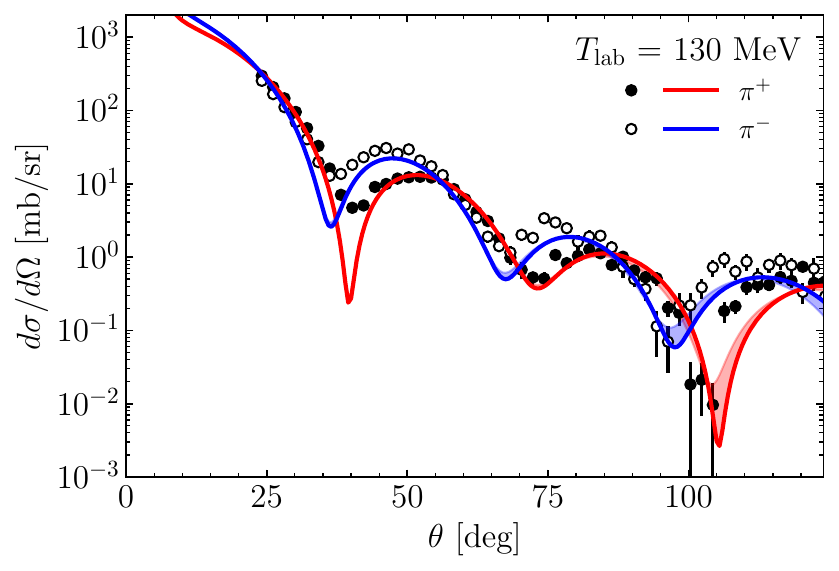}
		\includegraphics[width=0.495\textwidth]{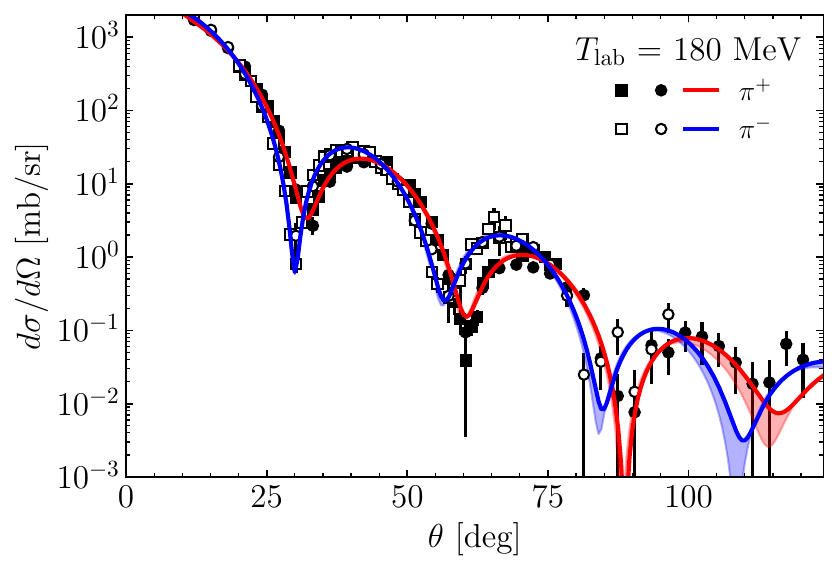}
		\hfill
		\includegraphics[width=0.495\textwidth]{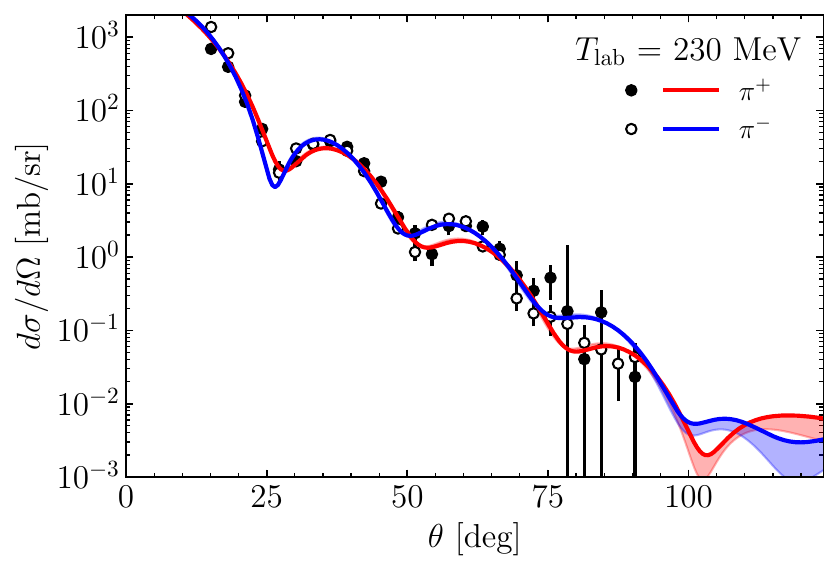}
		\caption{
			Predicted differential cross sections as a function of the scattering angle in the c.m. frame for $\pi^\pm$ elastic scattering on ${}^{48}$Ca using the full second-order potential. 
			Solid curves and closed markers stand for $\pi^+$; dashed and open markers for $\pi^-$. 
			The curves and markers have the same meaning as in Fig.~\ref{plt:48CaP_1st}.
			The curves are the same as in Figs.~\ref{plt:48CaP_1st} and~\ref{plt:48CaM_1st}, while the experimental data are renormalized according to Table~\ref{tabl:fit-N}.
		}
		\label{plt:48Ca}
	\end{figure*}

	Figures~\ref{plt:48CaP_1st} and~\ref{plt:48CaM_1st} illustrate the impact of the second-order component of the scattering potential on the differential cross section for positive and negative pion scattering off  ${}^{48}$Ca, respectively. 
	Theoretical predictions based on the derived second-order potential, i.e., the sum of Eqs.~(\ref{V(1)-fin}) and~(\ref{V(2)-fin}), are compared to those that incorporate only the first-order potential, Eq.~(\ref{V(1)-fin}).
	The results demonstrate that the second-order contribution, $V^{[2]}$, plays a significant role throughout the $\Delta$-resonance region.
	Although its impact is most pronounced at lower energies, it remains substantial even at pion laboratory kinetic energy $T_\text{lab} = \SI{230}{MeV}$.
	In particular, $V^{[2]}$ affects primarily the magnitude of the differential cross sections, with the diffraction minima shifting by only a few degrees compared to the first-order calculations.
	
	Figures~\ref{plt:48CaP_1st} and~\ref{plt:48CaM_1st} also explicitly illustrate how differential cross sections depend on the choice of the nuclear interaction. 
	The solid curves correspond to results obtained with $\Delta \text{NNLO}_{\text{GO}}(394)$, while the shaded bands indicate the difference between predictions from $\Delta \text{NNLO}_{\text{GO}}(394)$ and $\text{NNLO}_{\text{sat}}(450)$, thereby illustrating the sensitivity to the choice of interaction.
	Comparing the effects of these two interactions on pion-nucleus scattering is particularly informative, as both reproduce nuclear structure observables well but predict different neutron skin thicknesses,%
	\footnote{
		The neutron skin thickness is defined as the difference between the root-mean-square radii of the neutron and proton densities.
	}
	$r_{np}$, of ${}^{48}$Ca.
	Specifically, $\text{NNLO}_{\text{sat}}(450)$ yields $r_{np} = \SI{0.15\pm 0.01}{fm}$~\cite{Hagen2016}, while $\Delta\text{NNLO}_{\text{GO}}(394)$ gives $r_{np} = \SI{0.13\pm 0.01}{fm}$, where theoretical uncertainties account for the dependence on the truncation scheme (CCSD versus CCSDT-1) and the model space.
	As follows from the plots, the choice of interaction has minimal effect on the predictions obtained from the first-order potential, which depends only on the CC one-body densities. 
	In contrast, the uncertainty band associated with the second-order potential arises primarily from differences in the HF correlation functions predicted by the two interactions. 
	This uncertainty becomes more pronounced at larger angles, where the momentum transfer is higher, while remaining small at forward angles.
	Note that the uncertainty arising from the error matrix of the model parameters determined in Ref.~\cite{Tsaran:2023qjx} is considerably smaller than the uncertainty associated with the choice of nuclear interaction.
	This difference becomes especially pronounced at large scattering angles, where it can amount to an order of magnitude.

	Overall, the model demonstrates good agreement with experimental data, successfully reproducing the general features of the angular distributions.
	However, notable deviations with the data of Ref.~\cite{Gretillat:1981bq} remain at a pion laboratory kinetic energy of $\SI{130}{MeV}$, particularly for $\pi^-$ scattering.
	For $\pi^+$, the deviations are limited to an underestimation of the first diffraction minimum and a shift of approximately $5^\circ$ in the position of the third minimum. 
	In contrast, the $\pi^-$ cross section displays broader deviations across the full angular range.
	These deviations for $\pi^-$ are difficult to attribute to limitations of the model, as the lower-energy data at $\SI{116}{MeV}$, which are more sensitive to higher-order rescattering and Coulomb effects, are well described for both pion charges; the $\SI{180}{MeV}$ data are also accurately reproduced.
	Moreover, we have verified that $\pi^\pm$ scattering on ${}^{40}$Ca at $\SI{130}{MeV}$ is also rather well reproduced (see Fig.~\ref{plt:40Ca}). 
	Taken together, these observations suggest that the observed disagreement for negative pions at $\SI{130}{MeV}$ may stem from inconsistencies in the experimental data rather than a drawback in the theoretical framework.

	While the experimental error bars in Figs.~\ref{plt:48CaP_1st} and~\ref{plt:48CaM_1st} reflect statistical uncertainties, the differential cross-section data are also subject to various sources of systematic error, e.g., detector efficiencies, event data normalization, and background fitting methods. 
	To account for fully correlated systematic effects within each data set (instrumental errors), normalization parameters $N_i$ are introduced.
	The associated normalization uncertainties $\Delta N_i$, as reported in Refs.~\cite{Gretillat:1981bq} and~\cite{Boyer:1984bs}, are 6\% for each SIN $\pi^-$ data set, 9\% for $\pi^+$, and 15\% for the LAMPF data sets (these values are listed in the third and fifth columns of Table~\ref{tabl:fit-N}).

	\begin{table}[h]
		\caption{Fitted normalization parameters for SIN~\cite{Gretillat:1981bq} and  LAMPF~\cite{Boyer:1984bs} data. The first column indicates the pion energy, the third (fifth) column -- the experimental normalization uncertainties for $\pi^-$ ($\pi^+$) data, and the fourth (sixth) column -- the values of $\Delta N^\text{fit} = (N_i - 1)\, \%$ for $\pi^-$ ($\pi^+$) data obtained from the fit.}
		\begin{tabular}{cccccc}
			\hline\hline
			&  & \multicolumn{2}{c}{$\pi^-$} & \multicolumn{2}{c}{$\pi^+$} \\
			$T_\text{lab}$ [MeV] & Facility & $\Delta N_i \ [\%]$ & $\Delta N_i^\text{fit} \ [\%]$ & $\Delta N_i \ [\%]$ & $\Delta N_i^\text{fit} \ [\%]$  \\
			\hline
			116 & SIN   & $\pm 6$  &  $-1.1 \pm 4.6$ & $\pm 9$  &  $1.1 \pm 5.3$ \\
			130 & LAMPF & $\pm 15$ &  $-12.6 \pm 6.4$  & $\pm 15$  & $-3.7 \pm 4.8$  \\
			180 & LAMPF & $\pm 15$ &  $3.8 \pm 5.3$  & $\pm 15$  & $-15.2 \pm 3.1$ \\
			180 & SIN   & $\pm 6$  &  $1.4 \pm 4.3$  & $\pm 9$  & $0.7 \pm 4.5$ \\
			230 & LAMPF & $\pm 15$ &  $-10.0 \pm 7.5$ & $\pm 15$  &  $7.9 \pm 9.1$ \\
			\hline
		\end{tabular}
		\label{tabl:fit-N}
	\end{table}
	
	To test the consistency of our model with the experimental data while accounting for systematic uncertainties, we minimize a modified $\chi^2$  function defined as~\cite{DAgostini:1993arp}
	\begin{multline}
		\chi^2 = \sum_i \left[ \frac1{n_i} \sum_j^{n_i} \left( \frac{ d\sigma_j^{\text{Data}_i} - N_i^{-1}  d\sigma_j}{\Delta d\sigma_j^{\text{Data}_i}} \right)^2 
		+ \left( \frac{N_i - 1}{\Delta N_i} \right)^2
		\right],
		\label{chi-tot}
	\end{multline}
	with respect to the normalization factors $N_i$.
	Here, the index $i$ labels the individual datasets, each containing $n_i$ data points.
	The resulting best-fit normalization shifts, $N_i - 1$ lie within the $1\sigma$ intervals $\Delta N_i$ given by the data authors and are summarized in the fourth and sixth columns of Table~\ref{tabl:fit-N}.
	As shown in the table, the SIN data and our theoretical model are consistent, with normalization adjustments at the level of approximately 1\% for both 116 and $\SI{180}{MeV}$.
	In contrast, some of the LAMPF datasets require more substantial normalization shifts, still within the stated uncertainties, to achieve better agreement.
	For example, the $\SI{180}{MeV}$ $\pi^+$ data from LAMPF exhibit a shift of about 15\%.
	This adjustment brings the LAMPF data into agreement not only with the theoretical predictions but also with the corresponding SIN data.
	In this context, the 13\% normalization shift found for 130 MeV $\pi^-$, along with other shifts for LAMPF dataset, appears justified.

	The combined final results for the elastic scattering of both $\pi^{+}$ and $\pi^{-}$ on $^{48}$Ca, incorporating the normalization corrections to the data, are presented in Fig.~\ref{plt:48Ca}. 
	As seen in the figure, accounting for normalization uncertainties in the differential cross section data leads to a further improvement in the overall agreement between the model predictions and the experimental results.

	\section{Conclusion and outlook}
	\label{sec:conclusion}
	
	In this work, we have developed a second-order momentum-space potential for pion scattering on nuclei with non-zero isospin, extending the framework introduced in Ref.~\cite{Tsaran:2023qjx}. 
	Our approach provides a detailed treatment of intermediate charge-exchange processes in isospin-asymmetric nuclei by explicitly including $pp$, $nn$, and exchange $np$ correlations. 
	These second-order effects are expressed through corresponding two-nucleon correlation functions entering the second-order part of the potential.
	The nuclear structure inputs for the pion-${}^{48}$Ca potential were derived employing the state-of-the-art CC method for the proton and neutron one-body density and the HF approach for two-nucleon correlation functions.
	By employing different chiral EFT nuclear interactions, we were able to estimate the sensitivity of the scattering model to variations in the microscopic nuclear structure input.
	The pion-nucleon scattering amplitudes modified within the nuclear medium were adopted from the fit to $\pi^\pm$-${}^{12}$C scattering data presented in Ref.~\cite{Tsaran:2023qjx}.

	The developed model successfully reproduces the differential cross sections for $\pi^\pm$-${}^{48}$Ca elastic scattering in the $\Delta(1232)$-resonance energy region, with only mild sensitivity to the choice of the \textit{ab initio} nuclear interaction. 
	In particular, the correlation functions $C_{\tau_1 \tau_2, (\text{ex})}(\bm{q}_1, \bm{q}_2)$ derived within the HF framework using the $\Delta \text{NNLO}_{\text{GO}}(394)$ and $\text{NNLO}_{\text{sat}}(450)$ interactions exhibit good agreement, indicating their low sensitivity to the specific choice of interaction.
	
	The results obtained here for ${}^{48}$Ca corroborate our earlier findings in Ref.~\cite{Tsaran:2023qjx}, confirming that the inclusion of the second-order part of the potential is essential for achieving a consistent description of pion-nucleus scattering across the ${}^{12}$C-${}^{48}$Ca range in the $\Delta(1232)$ resonance energy region. 
	As the impact of the second-order contribution increases at lower energies, we anticipate that the choice of nuclear interaction will become a more significant source of theoretical uncertainty in the low-energy regime.

	As a next step, we plan to extend our analysis to low-energy pion scattering on calcium isotopes, aiming to refine the in-medium modifications of the isoscalar and isovector $s$-wave pion-nucleon amplitudes.
	Given the universality of our approach to both nuclear pion scattering and photoproduction processes (see Ref.~\cite{Tsaran:2024sue}), we expect the $C_{\tau_1 \tau_2, (\text{ex})}(\bm{q}_1, \bm{q}_2)$ correlation functions developed in this work to be directly applicable for computing second-order pion rescattering corrections in coherent pion photoproduction on ${}^{48}$Ca in order to compare with forthcoming data~\cite{WattsPC}.
	In future work, we also aim to investigate the sensitivity of coherent pion photoproduction and scattering observables to the neutron skin thickness of ${}^{48}$Ca.
	
	Finally, pion scattering and photoproduction on ${}^{40}$Ar are of particular interest for modern long-baseline neutrino oscillation experiments~\cite{MicroBooNE:2016pwy, DUNE:2016hlj}, where pion production is an important resonant process.
	Motivated by this, we furthermore plan to apply our formalism to ${}^{40}$Ar. 
	As a doubly open-shell nucleus, ${}^{40}$Ar presents additional challenges, which can be efficiently addressed using the double-charge-exchange equation-of-motion technique~\cite{Liu:2018mtu,Payne:2019wvy}, or through the recently developed deformed coupled-cluster framework~\cite{Hagen2022}.

	\section*{Acknowledgments}
	
	We express our gratitude to D. Watts for helpful discussions and C. L. Morris for sharing the LAMPF data.
	We thank Gaute Hagen for sharing the nuclear coupled cluster code developed at Oak Ridge (NUCCOR), as well as Joanna E.~Sobczyk and Weiguang Jiang for useful discussions.
	This work was supported by the Deutsche Forschungsgemeinschaft (DFG, German Research Foundation) through the SFB 1245 (Project-ID 279384907) and the Cluster of Excellence “Precision Physics, Fundamental Interactions, and Structure of Matter” (PRISMA+ EXC 2118/1, Project ID 39083149), ad by the U.S. Department of Energy, Office
	of Science, Office of Nuclear Physics, under the FRIB
	Theory Alliance award DE-SC0013617, and Office of
	Advanced Scientific Computing Research and Office of
	Nuclear Physics, Scientific Discovery through Advanced
	Computing (Sci-DAC) program (SciDAC-5 NUCLEI). This research used resources of the Oak Ridge Leadership Computing Facility located at Oak Ridge National Laboratory, which is supported by the Office of Science of the Department of Energy under contract No. DE-AC05-00OR22725. Computer time was provided by the Innovative and Novel Computational Impact on Theory and Experiment (INCITE) program and the supercomputer Mogon at Johannes Gutenberg Universit\"at Mainz.

	\vspace{0.5em}
	
	\appendix

	\section{One- and two-body density matrices}
	\label{sec:density-matrices-def}
	Density matrices offer a powerful framework for describing and analyzing many-body quantum mechanical systems, particularly atomic nuclei, covering the structural properties and correlations of nucleons within a nucleus~\cite{Lowdin:1955zzb}. 
	In this Section, we focus on one- and two-body density matrices of a system of $A$ nucleons in a nucleus.
	In coordinate space, the ground state of this system is described by a normalized antisymmetric $A$-body wave function $\Psi_0(x_1, \ldots, x_A)$, where $x_i = \{\bm r_i, s_i, \tau_i \}$ encompasses the position ($\bm r_i$), spin ($s_i$), and isospin ($\tau_i$) degrees of freedom of the $i$-th particle. 
	The one-body and two-body density matrices associated with the wave function $\Psi_0(x_1, \ldots, x_A)$ are defined in coordinate space as%
	\footnote{Note that the factor $A(A-1)$ in Eq.~(\ref{two-body-density-matrix-def}) is twice as large as that in another commonly used definition.
	}
	\begin{widetext}
		\begin{subequations}
			\begin{align}
				\rho_1(x_1; x_1^\prime) &\equiv A \mathlarger{ \int}  
				\left\{ \prod\limits_{i=2}^A \diff x_i \right\}
				\Psi_0^\dag(x_1, x_2, \ldots, x_A) \Psi_0(x_1^\prime, x_2, \ldots, x_A),
				\label{one-body-density-matrix-def}
				\\
				\rho_2( x_1,  x_2; x_1^\prime,  x_2^\prime)  &\equiv A (A - 1)
				\mathlarger{ \int} \left\{ \prod_{i=3}^A \diff x_i \right\} \Psi_0^\dag(x_1, \ldots, x_A) 
				\Psi_0(x_1^\prime, x_2^\prime, x_3, \ldots, x_A),
				\label{two-body-density-matrix-def}
			\end{align}
			\label{many-density-matrix-def}
		\end{subequations}
	\end{widetext}
	respectively. 
	To avoid ambiguity in the appendix, we use the subscripts in $\rho_1$ and $\rho_2$ to distinguish between one-body and two-body density distributions.
	The integration over $x_i$ in these equations and hereafter implies an integral over the corresponding spatial coordinates and a sum over the spin and isospin degrees of freedom
	\be
	\int \, \diff x_i \, \ldots = \sum_{s_i} \sum_{\tau_i} \int \, \diff \bm r_i  \, \ldots .
	\label{xi-integral-def}
	\ee
	Similarly, an arbitrary $n$-body density matrix is defined through an integral over the degrees of freedom of $A-n$ nucleons.

	The density matrix formalism allows for the straightforward derivation of the expectation values of any physical quantity $Q$ associated with a many-body system.
	In its most general form, the operator $\hat Q$ includes all possible interactions, from single particles to all particles interacting simultaneously, which can be represented as 
	\be
	\hat Q = \sum_{i=1}^A \hat Q^{(1)}_i + \sum_{i=1}^A \sum_{j \ne i}^A \hat Q^{(2)}_{ij} + \sum_{i=1}^A \sum_{j \ne i}^A \sum_{k \ne j}^A \hat Q^{(3)}_{ijk} + \cdots, 
	\ee
	where $\hat Q^{(n)}_i$ are operators acting on $n$ particles.
	The expectation value of $\hat Q$ can be expressed in terms of the density matrices as follows
	\begin{subequations}
		\begin{align}
			&\langle \Psi_0 | \hat Q | \Psi_0 \rangle = \bar Q_1 + \bar Q_2 + \ldots, \\
			&\bar Q_1 = \int \diff x_1  \diff x_1' \, \rho_1(x_1; x_1') \left\langle x_1' \left| \hat Q^{(1)} \right| x_1 \right\rangle ,\\
			&\bar Q_2 = \int \diff x_1 \diff x_2 \diff x_1' \diff x_2' \, \rho_2(x_1, x_2; x_1', x_2') \nonumber \\
			& \qquad\qquad\qquad\qquad\qquad
			\times \left\langle x_1', x_2' \left| \hat Q^{(2)} \right| x_1, x_2 \right\rangle,
			\label{bar_Q2}
		\end{align}
	\end{subequations}
	where the pattern continues similarly for higher-order terms involving more interacting particles.
	Thus, the density matrices defined by Eqs.~(\ref{many-density-matrix-def}) encapsulate the correlations and statistical properties of the nucleus, providing a way to compute the expectation values of nuclear one- and two-body operators.
	In this regard, they play a key role in deriving the potentials describing nuclear pion scattering and photoproduction.
	
	As follows from their definitions, the density matrices are Hermitian
	\be
	\begin{split}
		\rho_1(x; x^\prime) &= \rho_1^*(x^\prime; x), \\
		\rho_2(x_1, x_2; x_1^\prime, x_2^\prime) &=
		\rho_2^*(x_1^\prime, x_2^\prime; x_1, x_2).
	\end{split}
	\ee
	They are related by the formula
	\be
	\int \diff x_2 \, \rho_2(x_1, x_2; x_1^\prime, x_2) =
	(A - 1) \rho_1(x_1; x_1^\prime).
	\ee
	
	The diagonal elements of the density matrices, 
	\begin{subequations}
		\begin{align}
			\rho_1(x) &\equiv \rho_1(x; x), \\
			\rho_2(x_1, x_2) &\equiv
			\rho_2(x_1, x_2; x_1, x_2),
			\label{2body-distributions-def}
		\end{align}
		\label{dentsity-distributions-def}%
	\end{subequations}
	referred to as ``densities" or ``density distributions", are of special importance because they represent the probability densities of finding particles in specific states and positions.
	For an infinitesimal volume $\diff V_1$, $\diff V_1 \rho_1(x_1) / A$ provides the probability of finding a nucleon with spin $s_1$ and isospin $\tau_1$ in the volume $\diff V_1$ around the coordinate $\bm r_1$ when other nucleons have arbitrary positions, spins, and isospins. 
	Similarly,  $\rho_2(x_1, x_2) / A(A-1)$ gives the probability density for finding two nucleons with coordinates, spins, and isospins specified by $x_1$ and $x_2$.
	The density distributions, Eqs.~(\ref{dentsity-distributions-def}), are positive definite, and $\rho_2$ is symmetric in its coordinates: $\rho_2(x_1, x_2) = \rho_2(x_2, x_1)$.
	The normalization conditions for the densities are
	\begin{subequations}
		\begin{align}
			\int \diff x \, \rho_1(x) &= A, 
			\label{density-dist-norm-1} \\
			\int \diff x_1 \diff x_2  \,\rho_2(x_1, x_2) &= A(A-1).
		\end{align}
		\label{density-dist-norm}
	\end{subequations}
	Since nucleons obey Fermi-Dirac statistics, the total wave function of the nuclear system, $\Psi_0(x_1, \ldots, x_A)$, is antisymmetric with respect to the exchange of any two nucleons.
	This antisymmetry also applies to the density matrices
	\be
	\rho_2(x_1, x_2; x_1^\prime, x_2^\prime)  = 
	-\rho_2(x_2, x_1; x_1^\prime, x_2^\prime)  =
	-\rho_2(x_1, x_2; x_2^\prime, x_1^\prime).
	\ee

	\section{Density matrices for a Slater determinant state}
	\label{app:Slater}
	
	Here, we provide the evaluation of density matrices for a Slater determinant state, and Eqs.~\eqref{eq: D func correlation}--~\eqref{C(D)}.

	The one-body density matrix in coordinate space can be generally defined as the expectation value of the density operator
	\begin{align}
		&\hat{ \rho}_1(x,x^{\prime} ) = \hat{\psi}^{\dagger}(x) \hat{\psi}(x^{\prime}).
	\end{align}
	Correspondingly, the non-diagonal one-body density operators for protons and neutrons take the form
	\begin{align}
		& \hat{\rho}_{\tau}(\bm r, \bm r^{\prime}) = 
		\sum_{\sigma} \hat{\psi}^{\dagger}(\bm r \sigma \tau) \hat{\psi}(\bm r^{\prime} \sigma \tau).
	\end{align}
	
	While the general expression for the two-body density operator is given in Eq.~\eqref{two-body-density-matrix-(adag,a)}, for our purposes it is sufficient to consider only its spin-independent part (see Eq.~\eqref{2b-density-mom-sp}) defined as
	\begin{multline}
		\hat{\rho}_2(\bm r_1, \bm r_2; \tau_1 \tau_2; \tau_1^{\prime} \tau_2^{\prime})
		\\
		= \sum_{\sigma_1 \sigma_2  } \hat{\psi}^{\dagger}(\bm r_1 \sigma_1 \tau_1) \hat{\psi}^{\dagger}(\bm r_2 \sigma_2 \tau_2) \hat{\psi}(\bm r_2 \sigma_2 \tau_2^{\prime} ) \hat{\psi}(\bm r_1 \sigma_1 \tau_1^{\prime} ).
	\end{multline}

	We now take the expectation value of $\hat{\rho}_2$ on the HF ground state wave function $\ket{\Phi_0}$.
	By applying the Wick theorem, we arrive at 
	\begin{multline}
		\rho_2(\bm r_1, \bm r_2; \tau_1 \tau_2; \tau_1^{\prime} \tau_2^{\prime}) =  \mel{\Phi_0}{\hat{\rho}_{2}}{\Phi_0}   \\
		= \sum_{\sigma_1 \sigma_2  } \expval{\hat{\psi}^{\dagger}(\bm r_1 \sigma_1 \tau_1) \hat{\psi}(\bm r_1 \sigma_1 \tau_1^{\prime} )}
		\expval{\hat{\psi}^{\dagger}(\bm r_2 \sigma_2 \tau_2) \hat{\psi}(\bm r_2 \sigma_2 \tau_2^{\prime} )}  \\
		-\expval{ \hat{\psi}^{\dagger}(\bm r_1 \sigma_1 \tau_1) \hat{\psi}(\bm r_2 \sigma_2 \tau_2^{\prime}) }
		\expval{ \hat{\psi}^{\dagger}(\bm r_2 \sigma_2 \tau_2) \hat{\psi}(\bm r_1 \sigma_1 \tau_1^{\prime} ) } .
	\end{multline}
	Given the isospin and spin conservation, we employ the relation
	\be
	\expval{ \hat{\psi}^{\dagger}(\bm r_1 \sigma_1 \tau_1) \hat{\psi}(\bm r_2 \sigma_2 \tau_2) } =
	\frac{1}{2}
	\delta_{\tau_1 \tau_2} \delta_{\sigma_1 \sigma_2} \rho_{\tau_1}(\bm r_1, \bm r_2),
	\ee
	which leads to the following expression for the HF two-body density
	\begin{multline}
		\rho_2(\bm r_1, \bm r_2; \tau_1 \tau_2; \tau_1^{\prime} \tau_2^{\prime}) 
		= \rho_{\tau_1}(r_1) \rho_{\tau_2}(r_2)
		\delta_{ \tau_1 \tau_1^{\prime} }
		\delta_{ \tau_2 \tau_2^{\prime} } \\
		- \frac{1}{2} \rho_{\tau_1}(\bm r_1, \bm r_2) 
		\rho_{\tau_2}(\bm r_2, \bm r_1) \delta_{\tau_1 \tau_2^\prime} \delta_{\tau_2 \tau_1^\prime}.
		\label{rho2-SD-exp}
	\end{multline}

	This expression shows that for a Slater determinant form of the system wave function, $ \rho(\bm q_1, \bm q_2; \tau_1, \tau_2; \tau_1, \tau_2)$ can be expressed entirely as a function of the one-body off-diagonal matrix.
	Also, the structure of Eq.~\eqref{rho2-SD-exp} reveals that the two-body density naturally decomposes into an "isospin-diagonal" ($\tau_1 = \tau_1^{\prime}$, $\tau_2 = \tau_2^{\prime}$) and an "isospin-exchange" contribution ($\tau_1 = \tau_2^{\prime}$, $\tau_2 = \tau_1^{\prime}$).
	Using the $D$ correlation function, Eq.~\eqref{eq: D func correlation}, we find 
	\begin{align}
		& \rho_2(\bm r_1, \bm r_2; \tau \tau; \tau \tau) = \rho_{\tau}(r_1) \rho_{\tau}(r_2) - D_{\tau \tau}(\bm r_1, \bm r_2), \\
		& \rho_2(\bm r_1, \bm r_2; np; np) = \rho_n(r_1) \rho_p(r_2), \\
		& \rho_2(\bm r_1, \bm r_2; np; pn) = - D_{np}(\bm r_1, \bm r_2),
	\end{align}
	from which one can read off the relation Eqs.~\eqref{eq: C and D corr HF}.
	Note that the vanishing of $C_{np}$ is specific to HF correlation functions and may not hold in general.

	\begin{figure*}[!t]
		\includegraphics[width=0.5\textwidth]{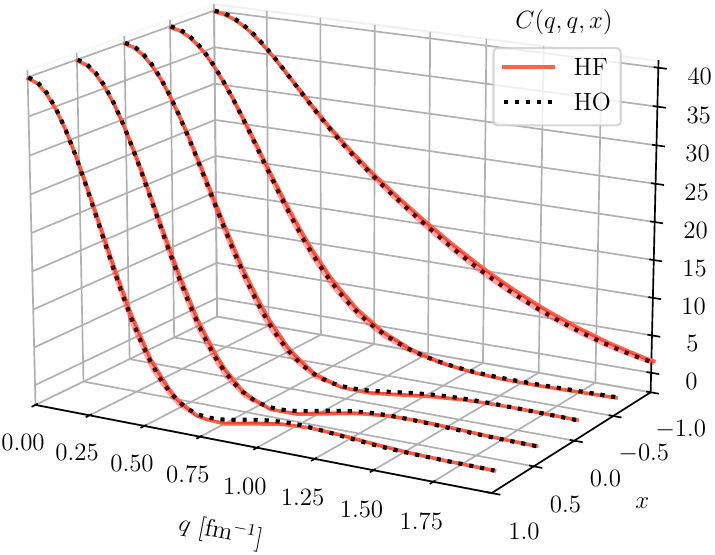}
		\hfill
		\includegraphics[width=0.49\textwidth]{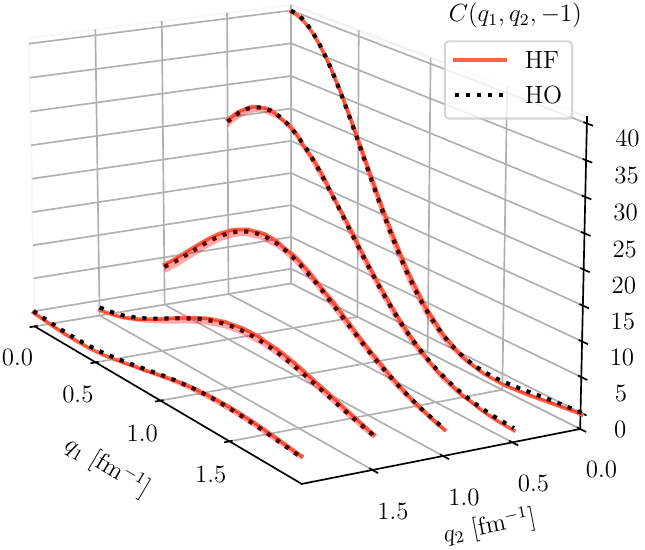}
		\caption{ 
			The correlation functions $C = C_{nn} + C_{pp}$ for ${}^{40}$Ca are shown for two momentum configurations: 
			$|\bm q_1| = |\bm q_2| = q$ and $x = \bm q_1 \cdot \bm q_2 / (q_1 q_2)$ (left panel) and $|\bm q_1| = q_1$, $|\bm q_2| = q_2$ for $x=-1$ (right panel).
			The solid red and dotted black curves are obtained using the HF wave function with $\Delta \rm{NNLO_{GO}}(394)$ and the non-interacting HO shell model used in Ref.~\cite{Tsaran:2023qjx}, respectively. 
			The error bars indicate HF results obtained with the $\rm{NNLO_{sat}}(450)$ interaction.
		}
		\label{fig:C_tot_40Ca}
	\end{figure*}

	\section{Application to ${}^{40}$Ca}
	\label{sec:40Ca}

	Here, we explore the applicability of the two-nucleon correlation functions developed in Sect.~\ref{sec: nuclear structure} to ${}^{40}$Ca directly employing the model parameters determined in Ref.~\cite{Tsaran:2023qjx}. 
	
	The three free parameters in our model effectively absorb, through the fitting procedure, various underlying processes and effects not explicitly included in the potential.
	For example, the imaginary part of the $\Delta$ self-energy is driven by true pion absorption processes.
	Similarly, while the correlation functions developed here and in~\cite{Tsaran:2023qjx} capture important nuclear-structure effects, it would be naive to expect them to account for all such contributions.
	In particular, accurate short-range correlations are not incorporated in either the HO or HF ground-state wave functions.
	Nonetheless, Ref.~\cite{Tsaran:2023qjx} demonstrates that a good description of the pion-nucleus scattering data can still be achieved. 
	
\begin{figure*}[!t]
		\includegraphics[width=0.5\textwidth]{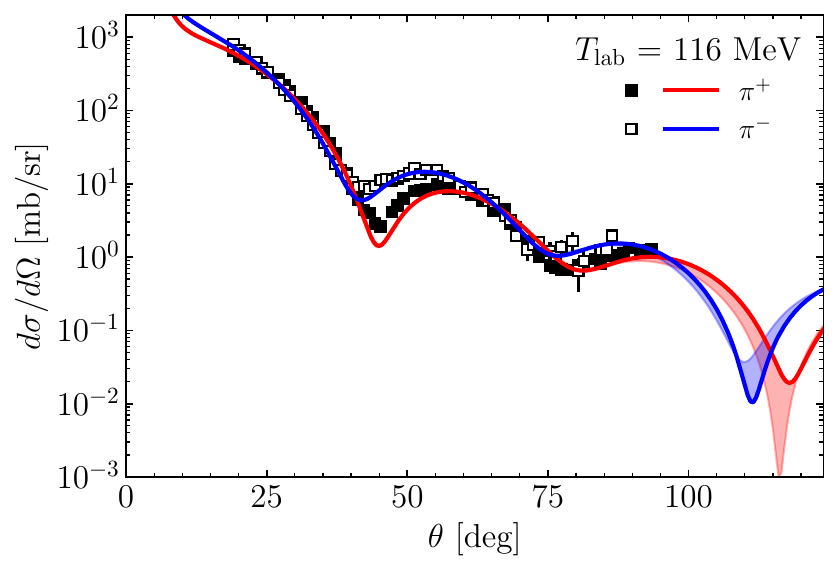}
		\hfill
		\includegraphics[width=0.49\textwidth]{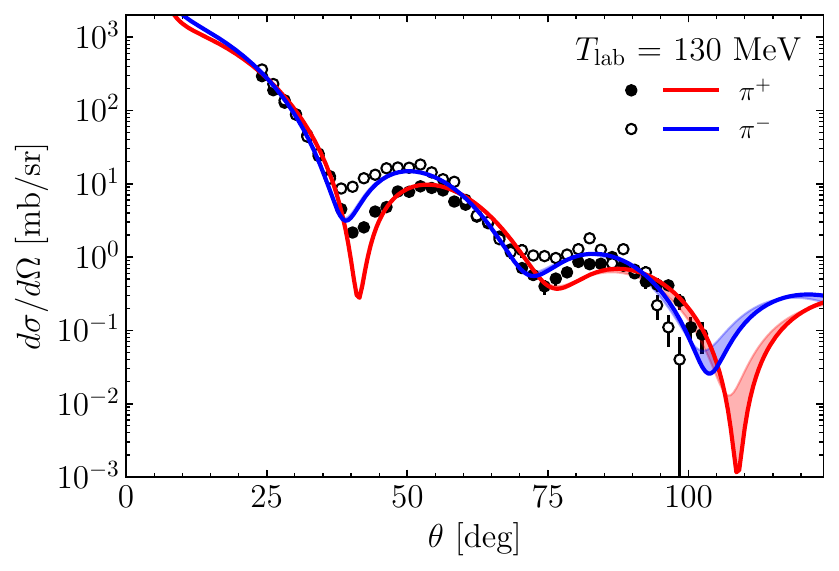} 
		\\
		\includegraphics[width=0.5\textwidth]{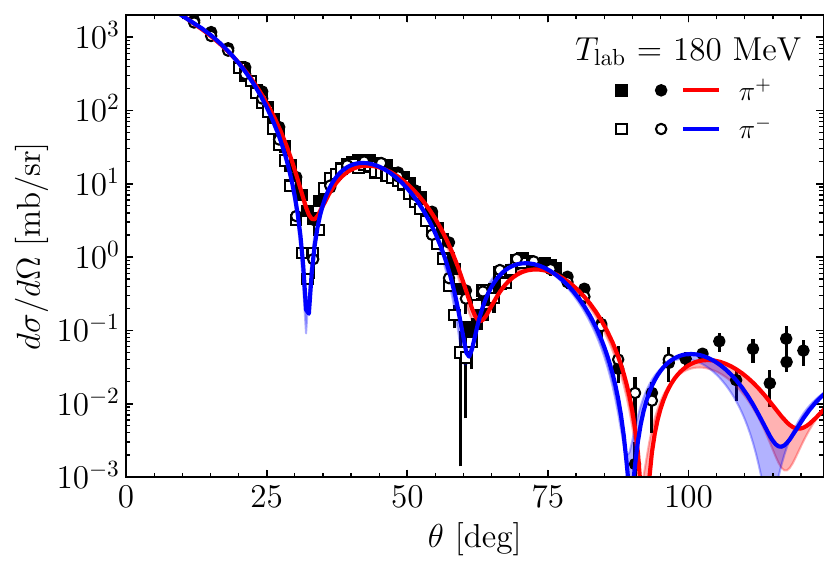}
		\hfill
		\includegraphics[width=0.49\textwidth]{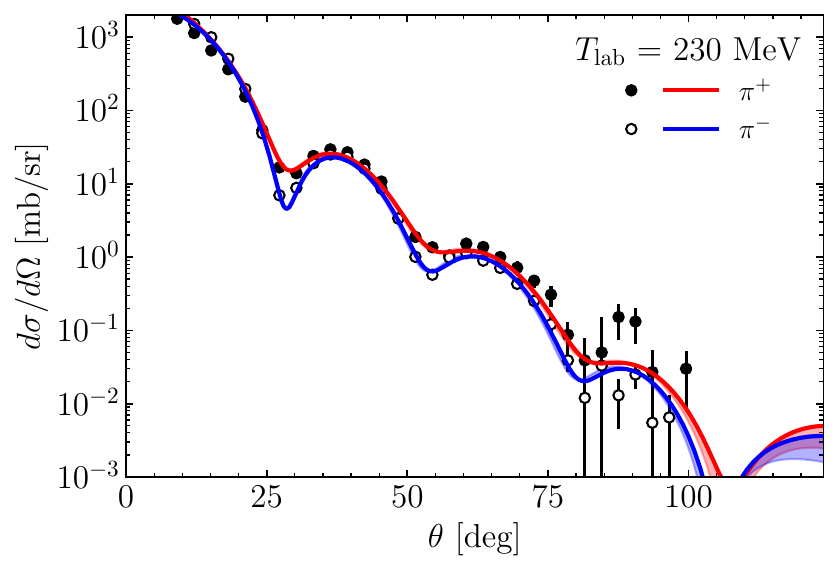}
		\caption{
			Predicted differential cross sections as functions of the scattering angle in the c.m. frame for $\pi^\pm$ elastic scattering off ${}^{40}$Ca using the full second-order potential. 
			Solid curves and closed markers stand for $\pi^+$; dashed and open markers for $\pi^-$. 
			The curves and markers have the same meaning as in Fig.~\ref{plt:48CaP_1st}.
		}
		\label{plt:40Ca}
	\end{figure*}

	The fitting procedure adjusts the model parameters to effectively account for missing short-range nucleon interactions and other omitted effects.
	For these reasons, before directly employing the parameters and amplitudes from Ref.~\cite{Tsaran:2023qjx}, it is essential to verify that the $C_{\tau_1 \tau_2, (\text{ex})}(\bm{q}_1, \bm{q}_2)$ correlation functions derived using non-interacting HO shell model and HF approaches are based on similar assumptions and include comparable physical effects.
	This comparison is shown in Fig.~\ref{fig:C_tot_40Ca}, where the agreement between the non-interacting HO shell model and HF correlation functions is clearly visible.
	This consistency supports the use of the amplitudes from Ref.~\cite{Tsaran:2023qjx} in conjunction with the HF-based approach developed in this work.
	Finally, in Fig.~\ref{plt:40Ca}, we present our predictions for differential cross sections of pion scattering off ${}^{40}$Ca, obtained using the method described in this work.

	\bibliography{bibliography.bib}

\end{document}